\documentclass{sig-alternate-10pt}

\usepackage{cite}

\usepackage{algorithm}
\usepackage{algorithmic}
\usepackage{fixltx2e}
\usepackage{amsmath}
\usepackage{float}
\usepackage{cite}
\usepackage{url}

\newcommand{\EE}{EE}

\newcommand{\ratiotag}{throughput/energy}
\newcommand{\FT}{FlexibleThr}

\usepackage{xr}
\usepackage{multirow}

\begin{document}

\title{Energy-Efficient Data Transfer Algorithms\\ for HTTP-Based Services}


\numberofauthors{2}
\author{
\alignauthor
Tevfik Kosar\\
       \affaddr{Computer Science and Engineering}\\
       \affaddr{University at Buffalo (SUNY)}\\
       \affaddr{Buffalo, New York 14260}\\
       \email{tkosar@buffalo.edu}
\alignauthor
Ismail Alan\\
       \affaddr{Computer Science and Engineering}\\
       \affaddr{University at Buffalo (SUNY)}\\
       \affaddr{Buffalo, New York 14260}\\
       \email{ialan@buffalo.edu}
}    

\maketitle
\begin{abstract}
According to recent statistics, more than 1 zettabytes of data is moved over the Internet annually, which consumes several terawatt hours of 
electricity, and costs billions of US dollars to the world economy. HTTP protocol is used in the majority of these data
transfers, accounting for 70\% of the global Internet traffic. We claim that HTTP transfers, and the services based on HTTP, can become more energy efficient
without any performance degradation by application-level tuning of certain protocol parameters. In this paper, we analyze several application-level parameters that affect the throughput and energy consumption in HTTP data transfers, such as the level of parallelism, concurrency, and pipelining. We introduce SLA-based algorithms which can decide the best combination of these parameters based on user-defined energy efficiency and performance criteria. Our experimental results show that up to 80\% energy savings can be achieved at the client and server hosts during HTTP data transfers and the end-to-end data throughput can be increased at the same time.
\end{abstract}

\section{Introduction}
\label{sec:introduction}

With the emergence of complex scientific applications, social media, video over IP, and more recently the trend for Internet of Things (IoT), the global data movement requirements have already exceeded the exabyte scale. It is estimated that, in 2017, more IP traffic will traverse global networks than all prior ``Internet years" combined. The global IP traffic will reach an annual rate of 1.4 zettabytes, which corresponds to nearly 1 billion DVDs of data transfer per day for the entire year. It is also estimated that around 20 billion devices will be connected to Internet during the same year~\cite{Cisco_2014}.

The annual electricity consumed by the global data movement is estimated to be more than 100 terawatt hours at the current rate, costing more than 20 billion US dollars per year~\cite{Gupta_2003, MINTS_2012, Mahadevan_2009, Cisco_2014, gao2012s}. This fact has resulted in considerable amount of work focusing on power management and energy efficiency in hardware and 
software systems~\cite{Brooks:2000:WFA:339647.339657, rawson2004mempower, zedlewski2003modeling, gurumurthi2002using, economou2006full,  fan2007power, rivoire2008comparison, koller2010wattapp, hasebe2010power, vrbsky2013decreasing, qureshi2009cutting} as well as on power-aware networking~\cite{Katz_2008, Mahadevan_2009, Greenberg_2009, Heller_2010, Goma_2011, fu2012frequency}. 
On the other hand, there has been little work focusing on saving data transfer energy at the end systems (sender and receiver nodes). 

Research shows that approximately 25\% of total electricity consumption during the end-to-end data transfers occur at the end-systems on a global (intercontinental) network, and this number goes up to 60\% on a nationwide network, and up to 90\% on a local area network~\cite{Alan2015}. This ratio depends on the number of network devices (i.e., routers, switches, hubs, etc.) between the sender and receiver nodes, and how much power each device consumes. On any of these networks, decreasing the end-system power consumption would result in significant energy savings considering exabytes of data are moved, and terawatts of electricity is consumed in worldwide data movement every year. 

In prior work, Alan et al~\cite{Kosar_jrnl14} analyzed the effects of different protocol parameters such as TCP pipelining, parallelism and concurrency levels on end-to-end throughput versus total energy consumption in the context of the GridFTP protocol~\cite{R_Allcock05, NDM_2012}, which is a fast, reliable and secure extension of FTP and widely used in the scientific computing community. They introduced three novel data transfer algorithms which achieve high data transfer throughput using GridFTP while keeping the energy consumption during the transfers at the minimal levels~\cite{Alan2015}.

In this paper, we analyze the performance versus energy consumption trade-offs in HTTP (Hypertext Transport Protocol), which is the de-facto transport protocol for Web services ranging from file sharing to media streaming. Studies analyzing the Internet traffic~\cite{Mirkovic2015, Czyz2014} show that in recent years HTTP holds the largest share and accounts for 70\% of this global Internet traffic. Recent studies argue that HTTP will become the narrow waist of the future Internet, meaning the vast majority of Internet traffic is expected to run over HTTP regardless of the underlying transport protocol~\cite{popa2010http}. 
We introduce service level agreement (SLA) based HTTP data transfer algorithms which can decide the best combination of these parameters based on user-set energy efficiency and performance criteria. To the best of our knowledge, this is the first work proposing energy-aware data transfer algorithms for high-performance HTTP data transfers with a quantification of the key practical tradeoffs between energy reduction and QoS/performance of data transfers impacting customer SLAs.

The presented SLA-based HTTP transfer algorithms let the end users to define their throughput and energy requirements. While keeping the quality of service (transfer throughput in this case) at the desired level, they keep the power consumption at the minimum possible level via tuning of the application-level transfer parameters. With the help of our SLA-based energy-efficient data transfer algorithms, the Internet service providers will be able to minimize the energy consumption during data transfers without compromising the SLA with the customer in terms of the promised performance level, but still execute the transfers with minimal energy levels given the requirements. Such a capability will be crucial for almost all big IT companies which provide cloud-hosted, web-based, or Internet-of-Things (IoT) related services. 
Our experimental results show that up to 80\% of energy savings can be achieved at the end systems during HTTP data transfers and the end-to-end throughput can be increased at the same time.

The rest of this paper is organized as follows. Section II presents background information on energy-aware tuning of HTTP and discusses the related work in this area. Section III analyzes the effects of different protocol parameters on HTTP performance as well as on end-system energy consumption. Section IV presents our SLA-based data transfer algorithms. Section V discusses the experimental results; Section VI studies the effect of these algorithms on network-power consumption; and Section VII concludes the paper.

\section{Energy-Aware HTTP Tuning}

One common way to increase the data transfer throughput at the application level is through the tuning of protocol parameters such as 
pipelining, parallelism, and concurrency.

{\em Pipelining} targets the problem of transferring a large number of small files over the network~\cite{TCP_Pipeline, farkas2002, Cluster_2015, TCC_2016}. In a regular TCP transfer, acknowledgement for the previous transfer is waited before starting the next transfer, which may cause a delay of more than one Round-Trip-Time (RTT) between individual transfers. With pipelining, multiple unacknowledged transfers can be active in the network pipe at any time and the delay between individual transfers is minimized.

\begin{table*}[t]
\begin{center}
\caption{Characteristics of the datasets used in the analysis.} \label{tab:dataset-spec}
\begin{tabular}{ l*{6}{c}r}

{\bf Dataset} &{\bf Total Size} & {\bf Number of Files } & {\bf File Format } & {\bf Avg. File Size }  & {\bf Min - Max }  & {\bf Std. Dev.}  \\
\hline
HTML &  495 MB & 5000 & HTML & 102 KB & 50 KB - 150 KB & 29.14 KB   \\
Image  & 1258 MB & 500 & JPEG & 2.4 MB & 2 MB - 3 MB & 0.28 MB \\ 
Video  & 4458 MB & 20 & AVI & 222 MB & 202 MB - 249 MB &  14.78 MB  \\  
\end{tabular}
\end{center}
\vspace{-6mm}
\end{table*}

{\em Parallelism} sends different chunks of the same file using different data channels (i.e., TCP streams) at the same time 
and achieves high throughput by mimicking the behavior of individual streams and getting a higher share of the available bandwidth~\cite{R_Sivakumar00, R_Lee01, R_Balak98, R_Hacker05, R_Eggert00, R_Karrer06, R_Lu05, DADC_2008, DADC_2009, NDM_2011}. On the other hand, using too many simultaneous connections congests the network and the throughput starts dropping down. Predicting the optimal parallel stream number for a specific setting is a very challenging problem due to the dynamic nature of the interfering background traffic. 

{\em Concurrency} refers to sending multiple files simultaneously through the network using different data channels at the same time.
Most studies in this area do not take the data size and the network characteristics into consideration when setting the concurrency level~\cite{kosar04, Thesis_2005, Kosar09, JGrid_2012}. Liu et al.~\cite{R_Liu10} adapt the concurrency level based on the changes in the network traffic, but do not take into account other bottlenecks that can occur on the end systems. 

When used wisely, these techniques have a potential to improve the transfer performance at a great extent, but improper use of these parameters can also hurt the performance of the data transfers due to increased load at the end systems and congested links in the network. For this reason, it is crucial to find the best combination for these parameters with the least intrusion and overhead to the system resource utilization and power consumption. 

Prior work on application level tuning of transfer parameters mostly proposed static or non-scalable solutions to the problem with some predefined values for the subset of the problem space~\cite{globusonline, R_Hacker02, R_Crowcroft98, R_Dinda05}. The main problem with such solutions is that they do not consider the dynamic nature of the network links and the background traffic in the intermediate nodes. Managed File Transfer (MFT) systems were proposed which used a subset of these parameters in an effort to improve the end-to-end data transfer throughput~\cite{WORLDS_2004, ScienceCloud_2013, globusonline, Royal_2011, IGI_2012}.

Kasparan et al.~\cite{kaspar2010using} analyzed how HTTP pipelining affects throughput in wireless local area networks (WLAN) and high-speed downlink packet access (HSDPA) networks. Results showed enabling pipelining eliminates the waiting time between successive requests and effectively increases the aggregated throughput. The fine-grained segmentation of data and using pipelining to fill the bandwidth-delay product (BDP) results in a natural adaption to network heterogeneity and increases interruption-free data transfers. 

Natarajan et al.~\cite{natarajan2008multistreamed,natarajan2009multiple} showed that using a single HTTP stream for transferring independent web objects is very inefficient in high latency networks. Using concurrent HTTP channels for delivering such objects increases download rates and decreases browser response times by enabling concurrent rendering. Their work showed that the performance of using aggressive number of parallel HTTP flows depends on several factors such as available bandwidth, path latency, HTTP response size, and network congestion. 

Robert et al.~\cite{kuschnig2010improving} used parallel HTTP requests to increase usage of available bandwidth during Internet video streaming. They analyzed video streaming performance in terms of video quality at various packet loss rates. Their results showed that at high packet loss rates using multiple HTTP channels with large sized chunks perform much better than single HTTP channel without compromising TCP-friendliness. 
Li et al.~\cite{liu2011parallel} presented a parallel HTTP streaming method instead of traditional adaptive sequential streaming. In distributed network environments fetching media segments by parallel channels increases streaming performance and also copes with inefficient usage of resources.

None of the existing work in this area proposed any energy-aware data transfer algorithms for high-performance HTTP data transfers with a quantification of the key practical tradeoffs between energy reduction and end-to-end performance of data transfers impacting customer SLAs.

\section{Analysis of Protocol Parameter Effects}

Measuring the energy consumption of data transfers is a challenging task due to lack of hardware power monitoring options for most of the devices. Even though one can measure power consumption of a specific device using a power meter, it is not always possible to hook up power meters to all systems involved in a particular study, especially when they are remote servers operated by other entities.

In this work, we used two power models similar to those presented by Alan et al.~\cite{Kosar_jrnl14} to estimate the server power consumption during data transfers for two different cases of access privileges: {\em (i)} the fine-grained power model requires access to utilization information of four system components: CPU, memory, disk and NIC; {\em (ii)} the CPU-based  power model only needs utilization information of the CPU in the target system. Our approach resembles Mantis~\cite{economou2006full} in predicting power consumption of data transfers based on operating system metrics. It is non-intrusive, models the full-system power consumption, and provides real-time power prediction. It  requires a one-time model-building phase to extract power consumption characteristics of the system components. For each system component (i.e. CPU, memory, disk and NIC), we measure the power consumption values for varying load levels. Then, linear regression is applied to derive the coefficients for each component metric. The derived coefficients are used in our power model to predict the total power consumption of the data transfers.

These power models were used to analyze the power consumption of HTTP transfers at different levels of pipelining (pp), parallelism (p), and concurrency (cc) levels. Three different representative datasets were used during experiments in order to capture the throughput and power consumption differences based on the dataset type: {\em (i)} the HTML dataset is a set of raw HTML files from the Common Crawl project ~\cite{commoncrawl}; {\em (ii)} the image dataset is a set of photo files from~Flicker~\cite{flicker}; and {\em (i)} the video dataset is a set of video file from Jiku~\cite{jiku}. The details of these datasets are presented in Table~\ref{tab:dataset-spec}. 

\begin{figure}[t]
\begin{center}
\includegraphics[keepaspectratio=true,angle=0,width=85mm]{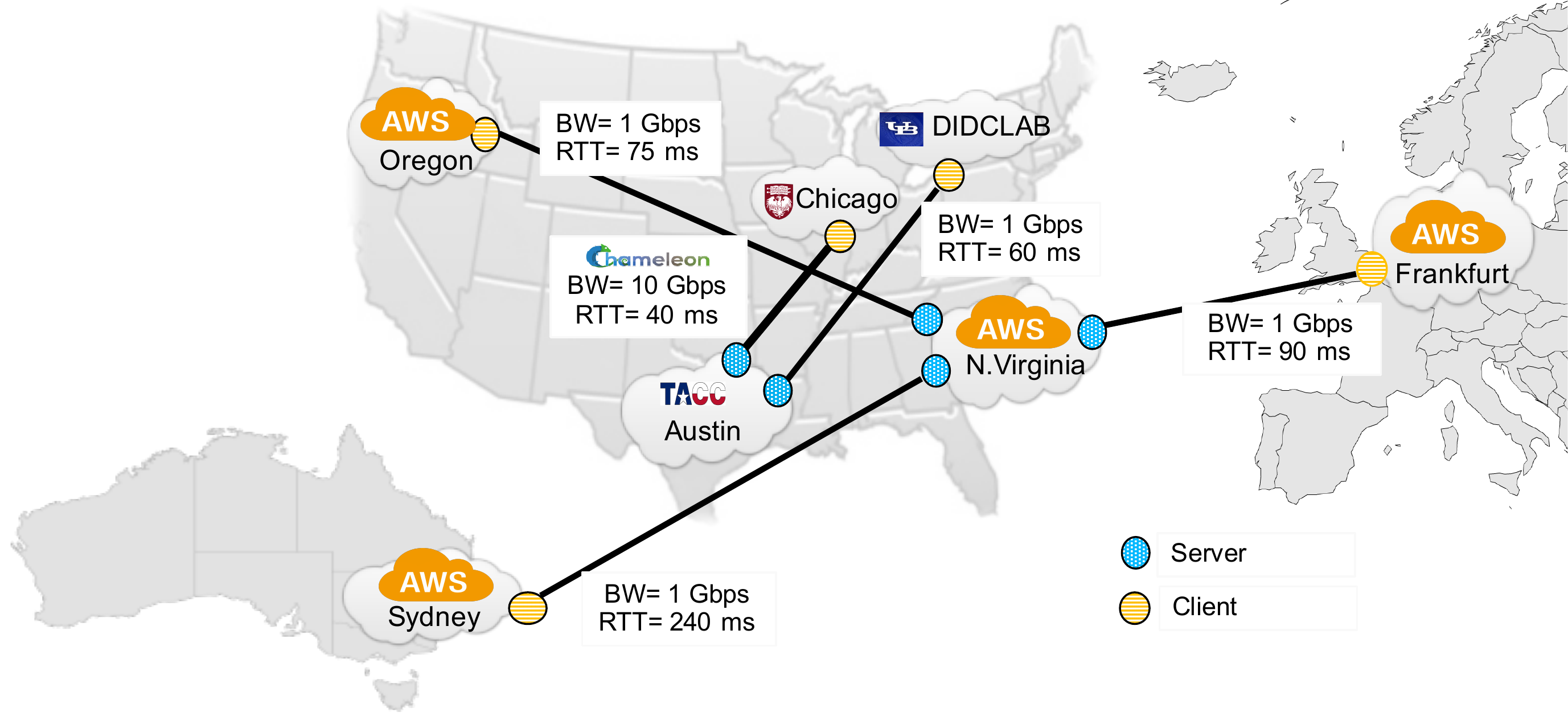}
\caption{Network map and specifications of the test environment.}
\label{fig:netmap}
\end{center}
\vspace{-3mm}
\end{figure}

Apache HTTP server~\cite{fielding1997apache} with default configuration and a custom HTTP client based on Apache HTTP Component libraries (which enables modification of protocol parameters) are used in the experiments. We also compared our client performance to two widely used command-line HTTP clients, wget~\cite{wget} and curl~\cite{curl}. Unfortunately these clients do not allow tuning of the protocol parameters. The experiments are conducted using two different server locations, five different client locations, and five different network settings. One of the servers is located at the Chameleon Cloud ~\cite{chamelon} node in Austin, Texas (USA) and it is serving the clients at the DIDCLab~\cite{DIDCLab} in Buffalo, New York (USA) and at the University of Chicago, Illinois (USA). The second server is located at the Amazon Web Services (AWS) Elastic Compute Cloud (EC2)~\cite{ec2} node in North Virginia (USA) and it serving clients located at the AWS nodes in Oregon (USA), Frankfurt (Germany), and Sydney (Australia). The locations and specifications of the used servers and clients as well as the capacities of the links between them are presented in Figure~\ref{fig:netmap}.

\begin{figure*}[t]
\begin{centering}
\begin{tabular}{cc}
 	\includegraphics[keepaspectratio=true,angle=0,width=55mm]{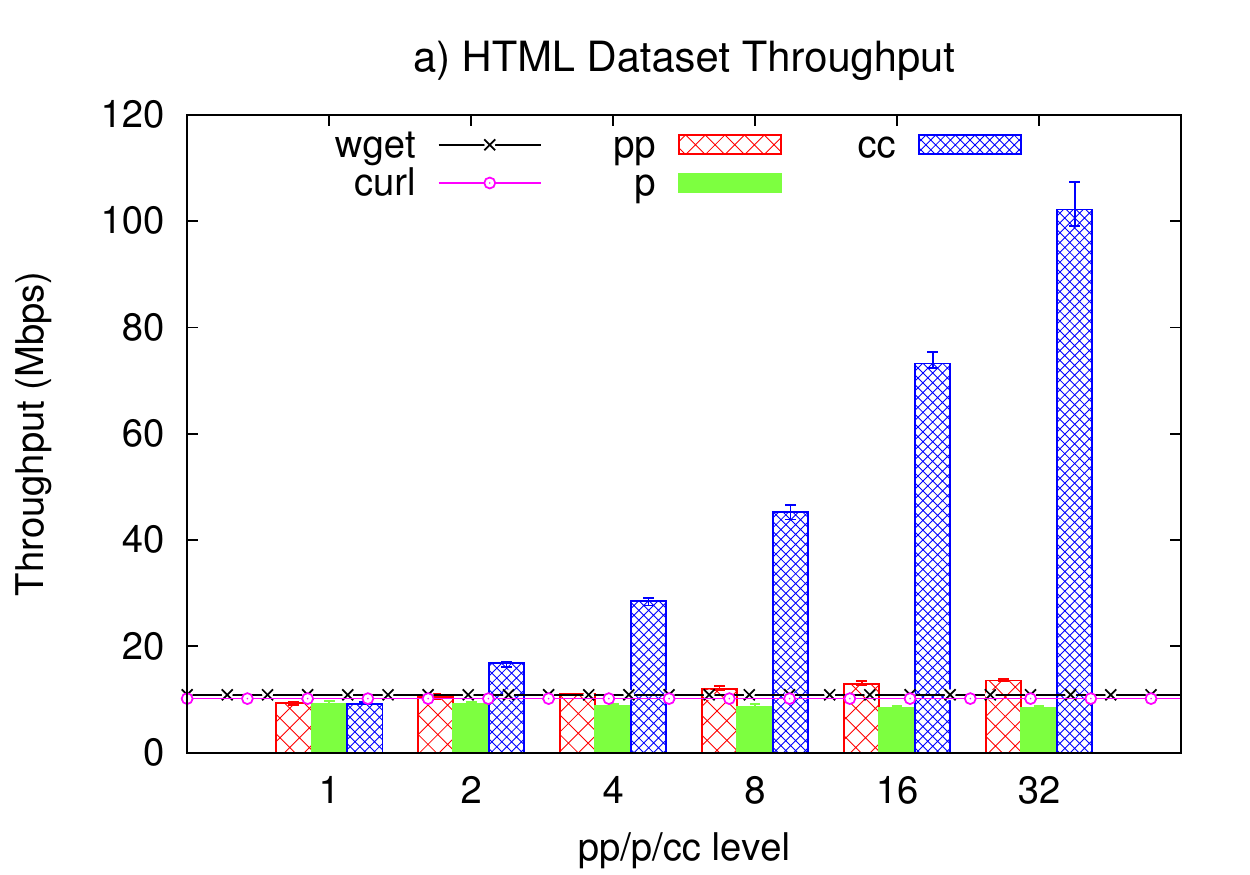}
 	\includegraphics[keepaspectratio=true,angle=0,width=55mm]{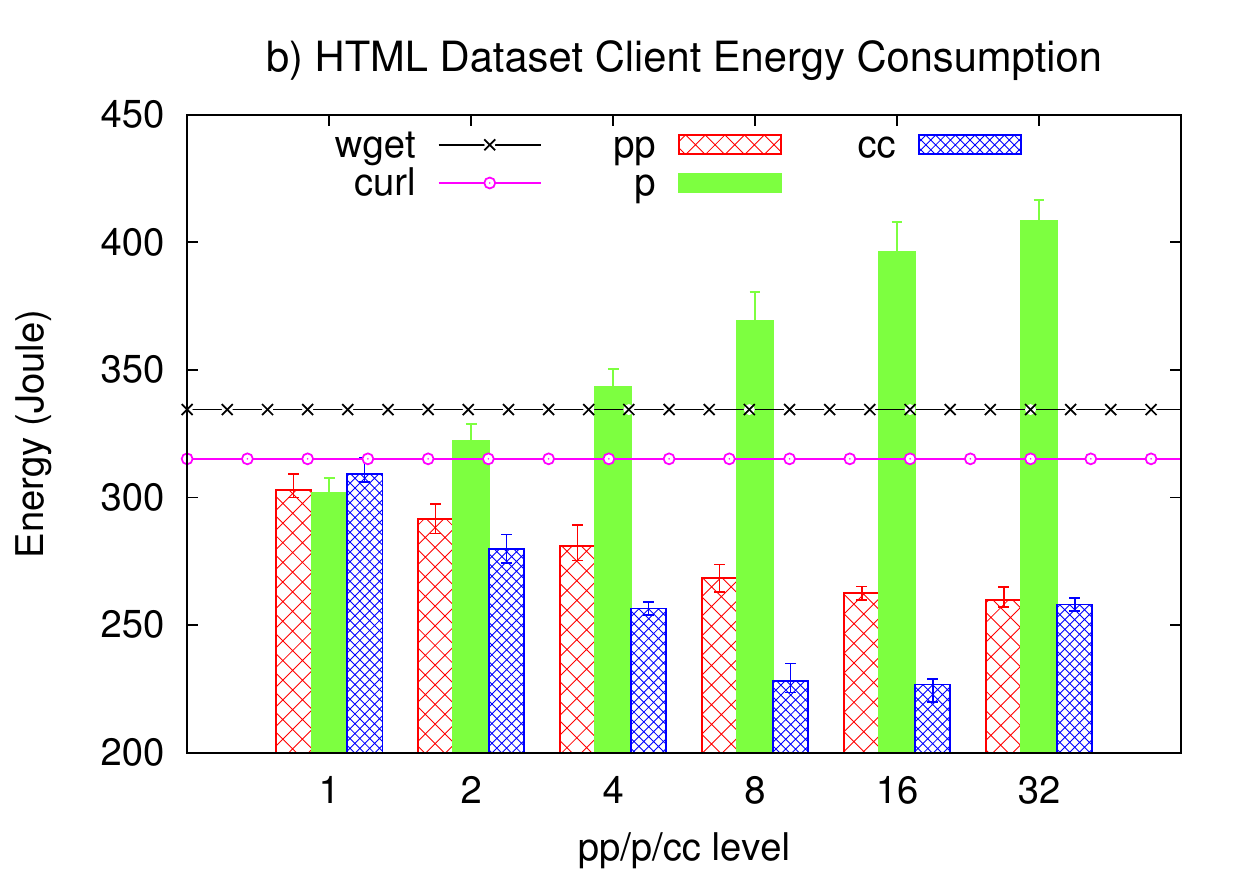}
 	\includegraphics[keepaspectratio=true,angle=0,width=55mm]{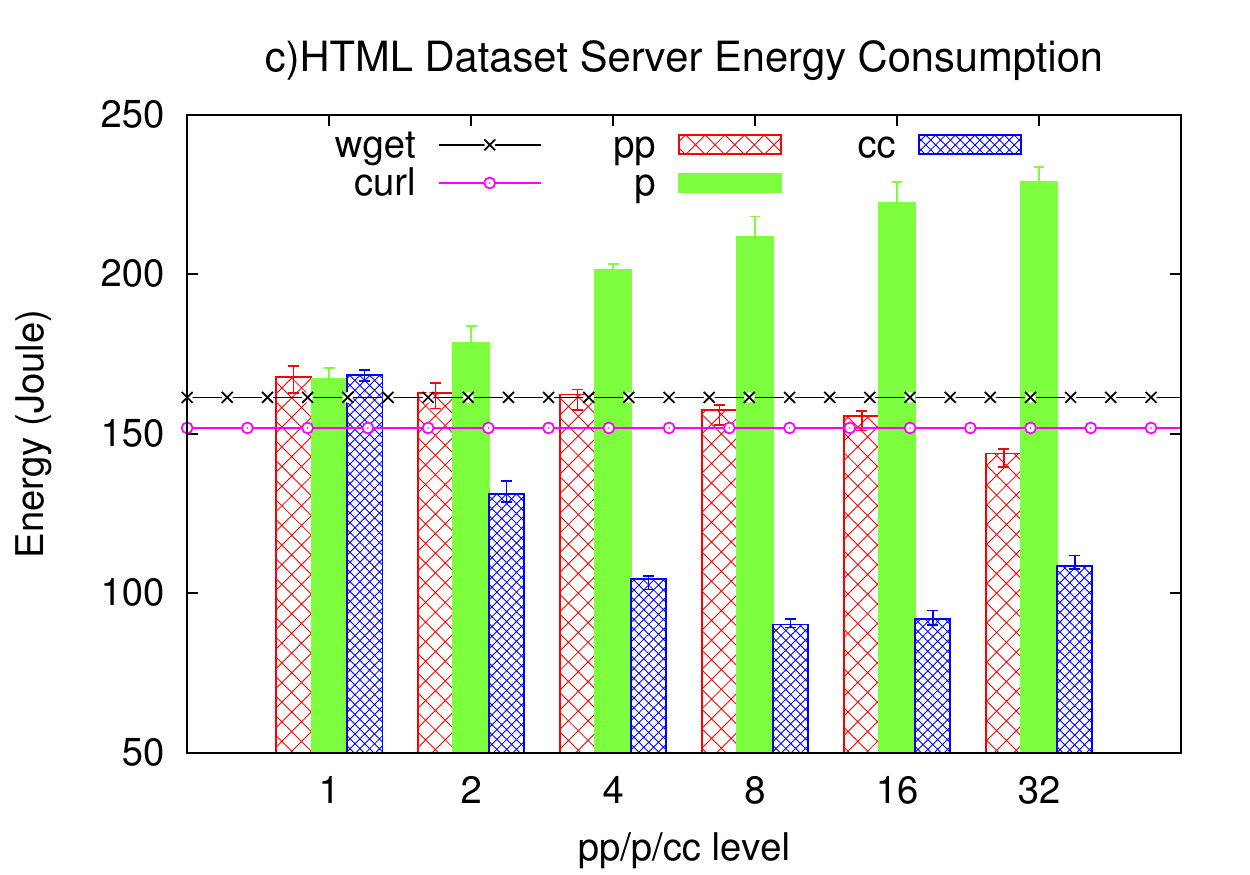}\\
	\includegraphics[keepaspectratio=true,angle=0,width=55mm]{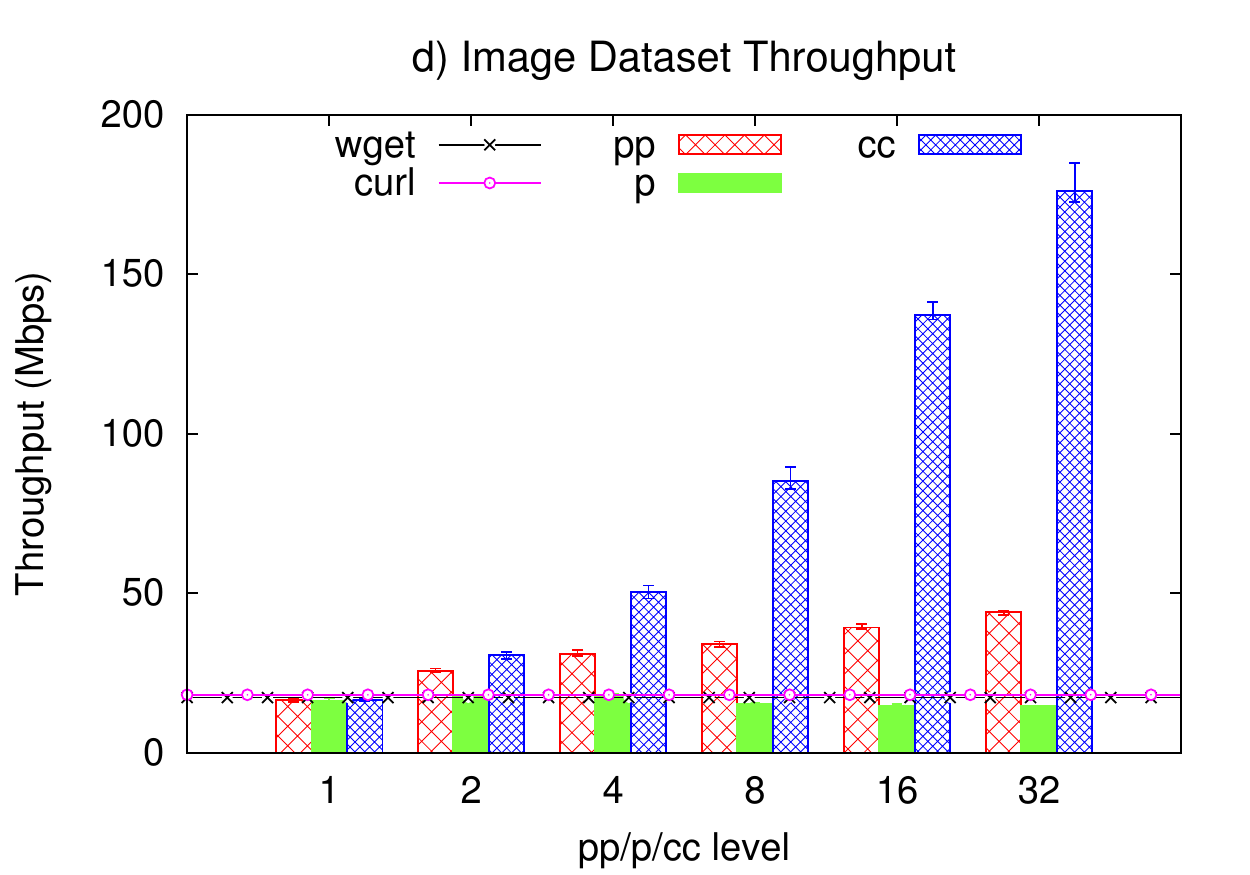}
 	\includegraphics[keepaspectratio=true,angle=0,width=55mm]{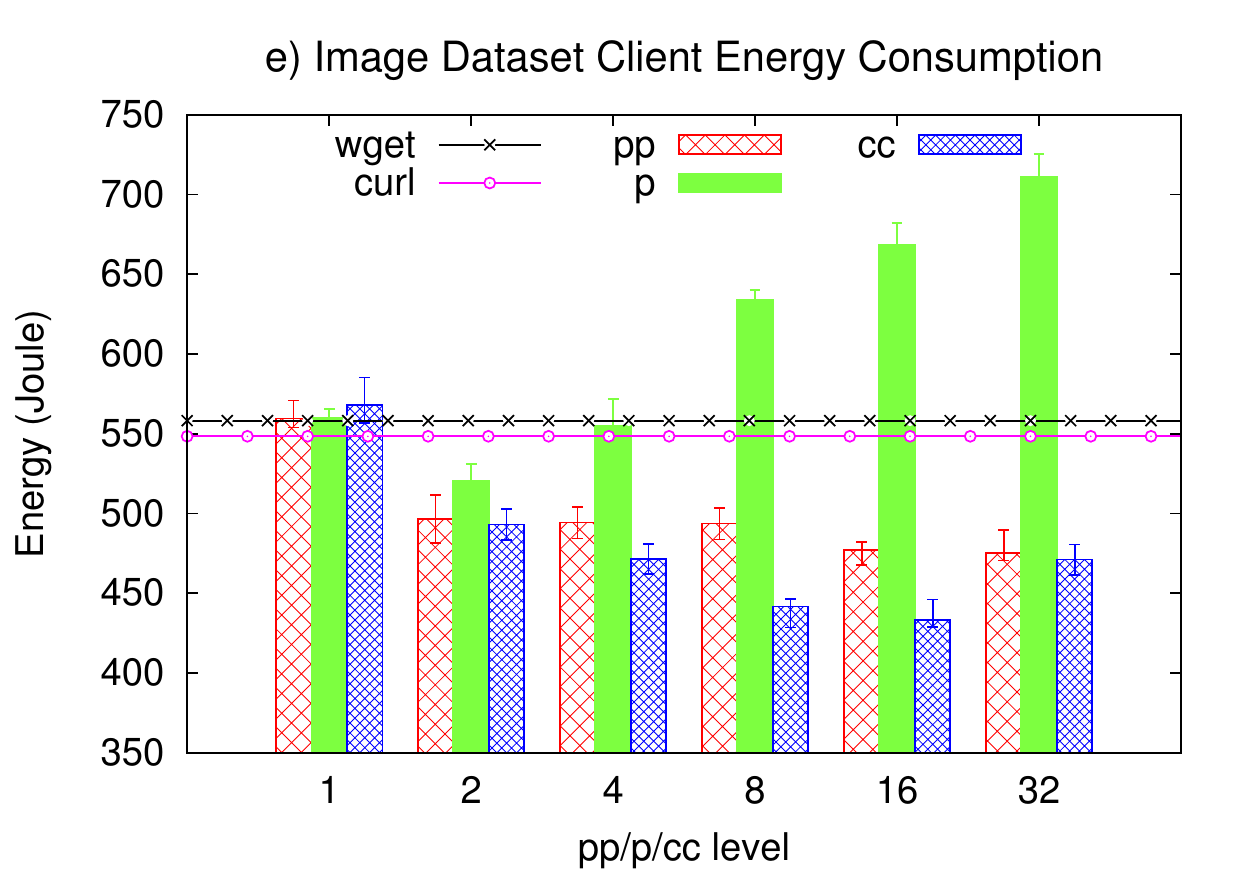}
 	\includegraphics[keepaspectratio=true,angle=0,width=55mm]{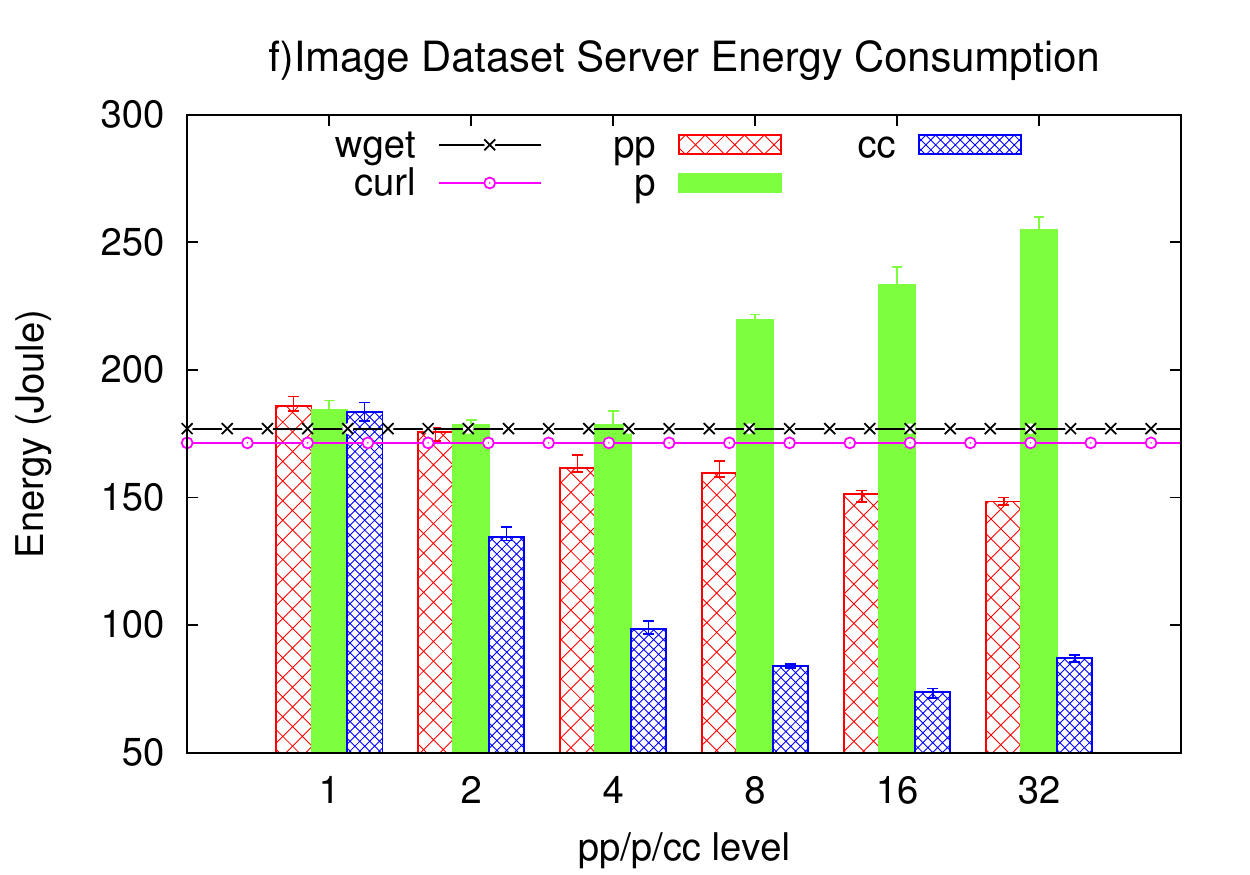}\\
 	\includegraphics[keepaspectratio=true,angle=0,width=55mm]{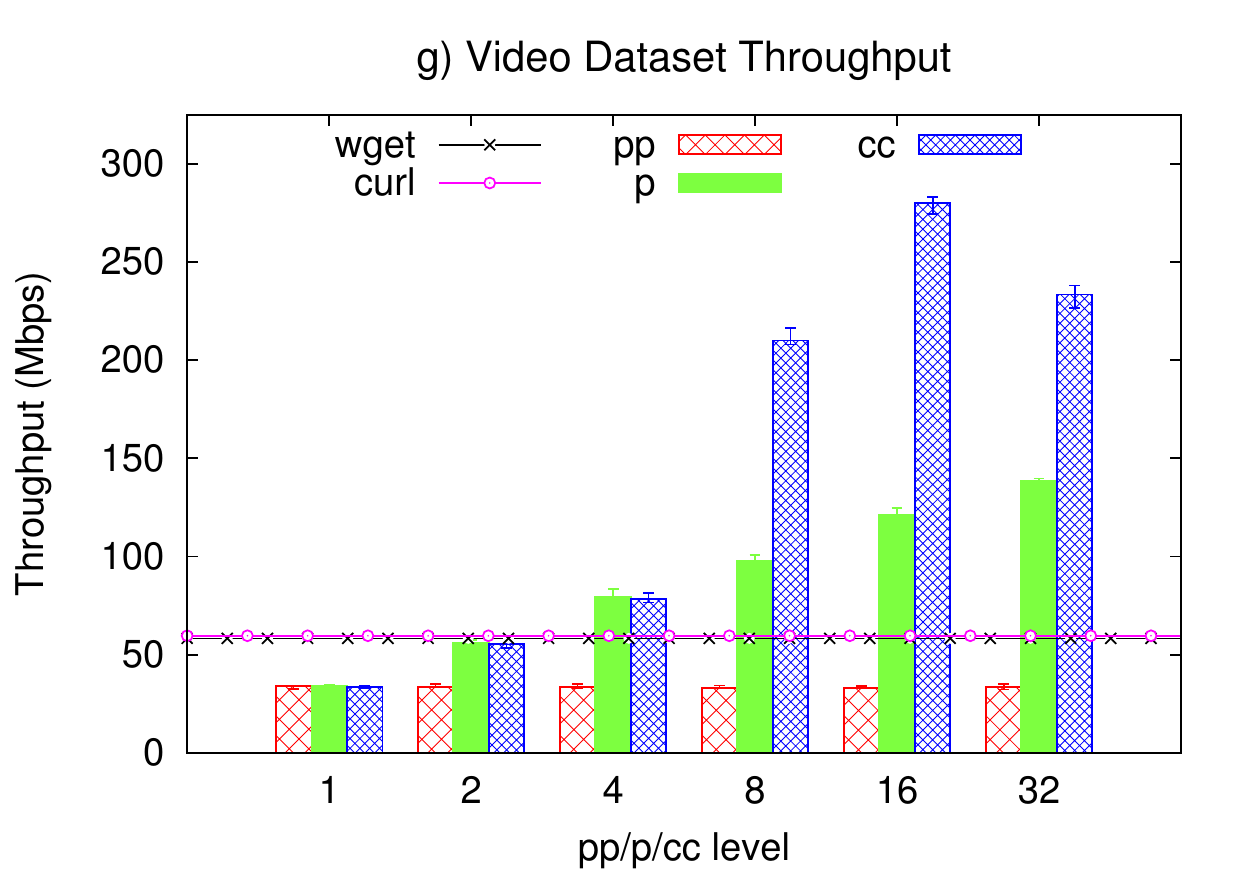}
 	\includegraphics[keepaspectratio=true,angle=0,width=55mm]{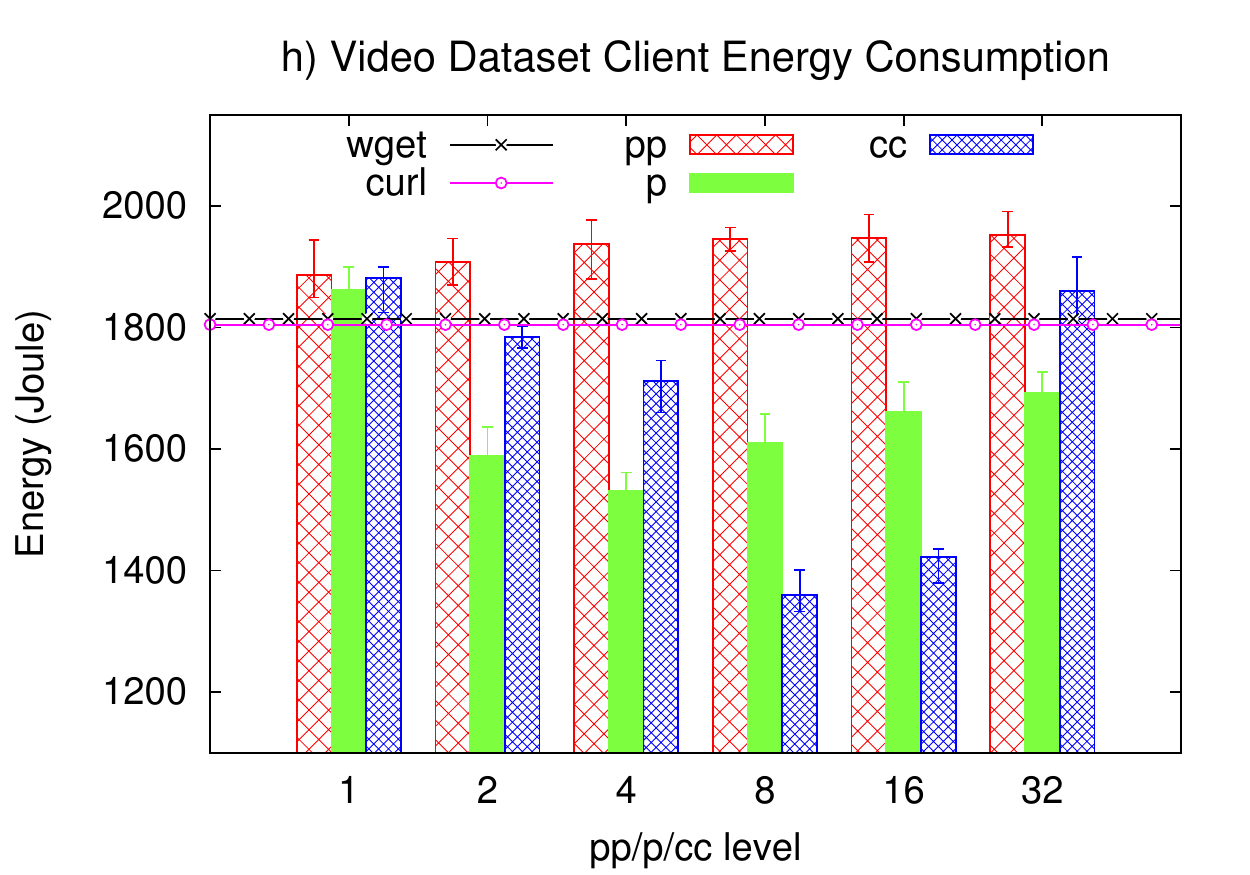}
 	\includegraphics[keepaspectratio=true,angle=0,width=55mm]{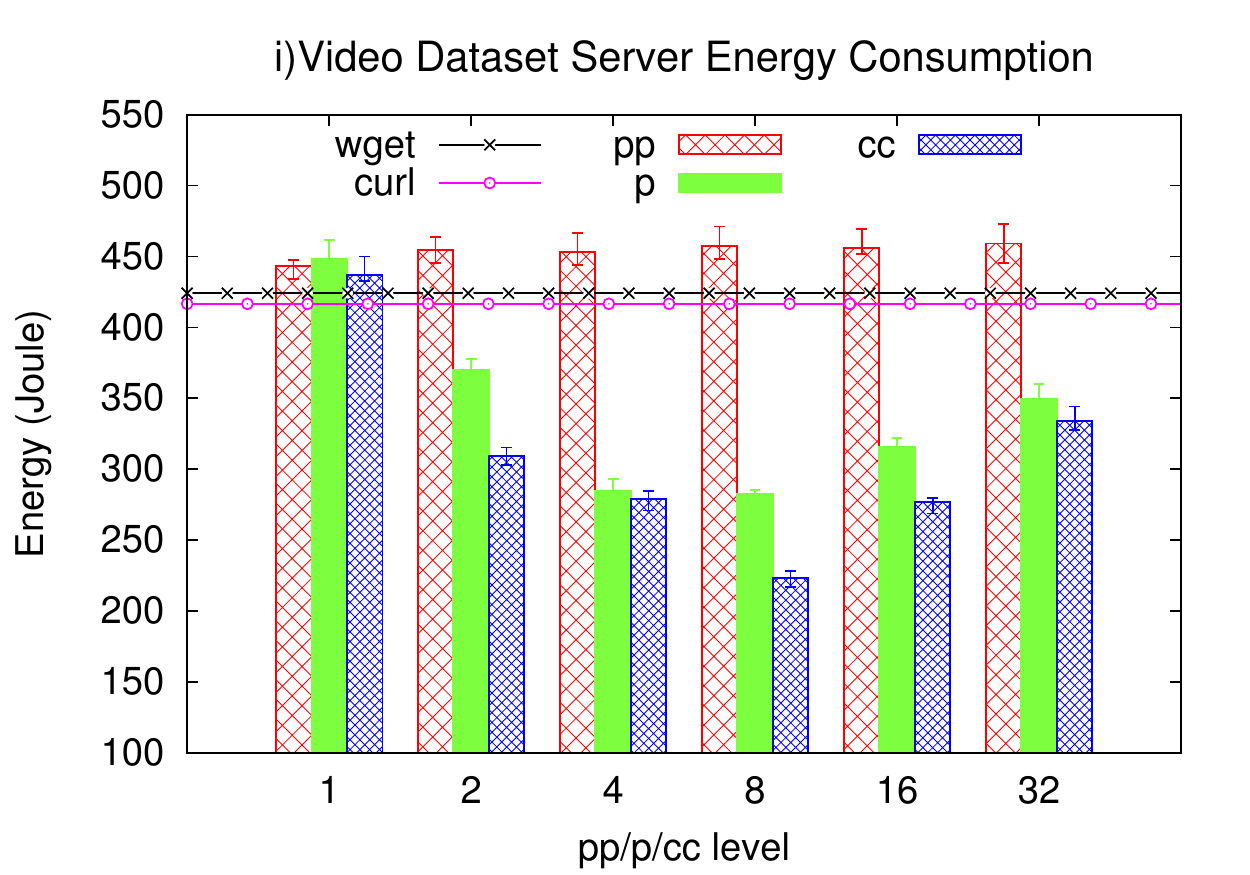}\\
\end{tabular} 
\caption{Performance vs energy consumption trade-offs of HTTP transfers from a web server in Austin to a client in Buffalo.} \label{fig:didclab-alamo}
\end{centering}
\end{figure*}

Figure~\ref{fig:didclab-alamo} presents the results of the data transfers between a web server at the Chameleon Cloud node in Austin, TX and the web client at DIDCLab in Buffalo, NY. The network bandwidth between these two nodes is 1 Gbps and the round trip time is 60 ms. We transferred each dataset only changing one parameter (i.e. pipelining, parallelism, or concurrency) at a time to observe the individual effect of each one on the end-to-end throughput versus the end-system (client and server) energy consumption. 
Energy consumption of a data transfer is calculated using the two power models we mentioned at the beginning of this section. We repeated each experiment at least five times at various times and took average of them. Since there is no dedicated network between the source and destination servers, and a shared internet connection is utilized, it is hard to achieve the maximum attainable network bandwidth. However, our results showed that we can achieve up to 280 Mbps throughput for the large (video) dataset. 

\begin{figure*}[t]
\begin{centering}
\begin{tabular}{cc}
 	\includegraphics[keepaspectratio=true,angle=0,width=55mm]{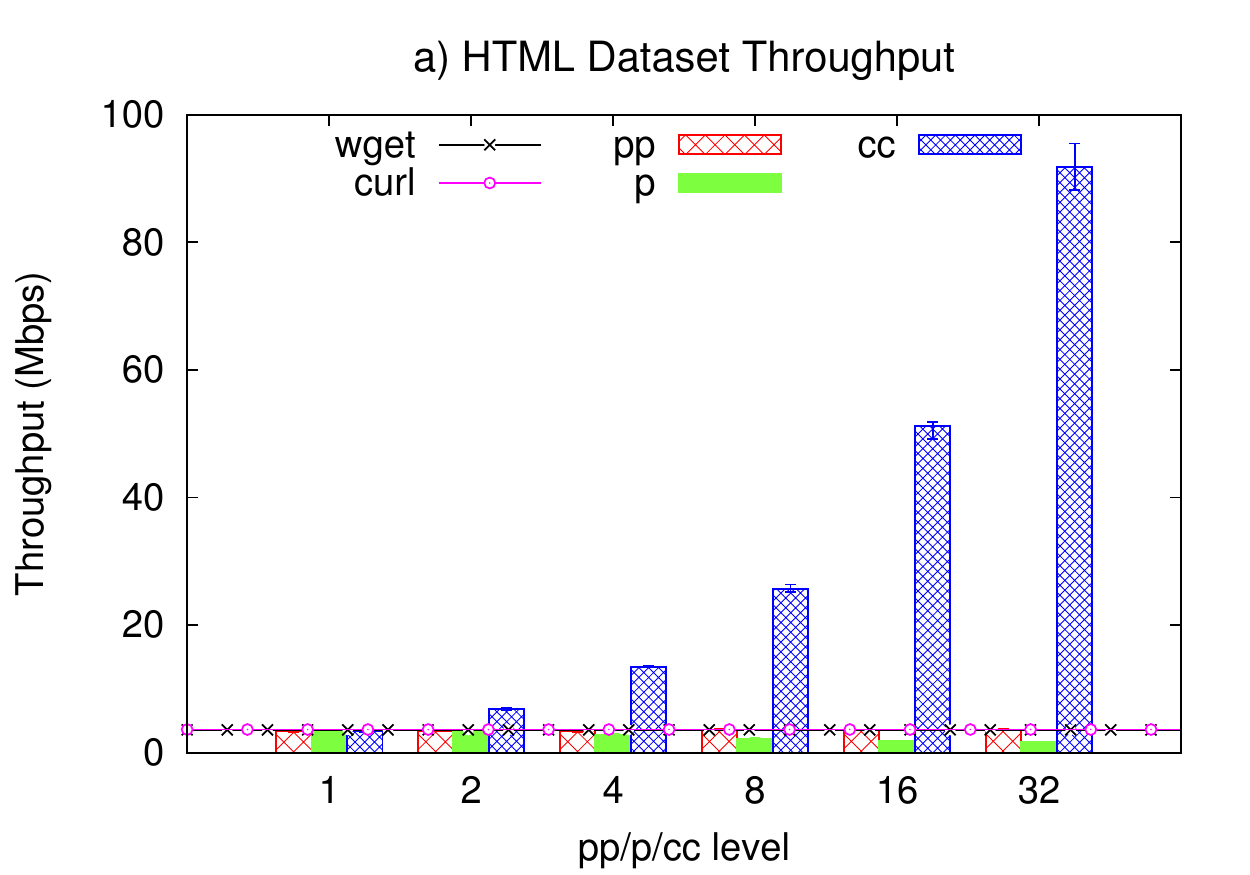}
 	\includegraphics[keepaspectratio=true,angle=0,width=55mm]{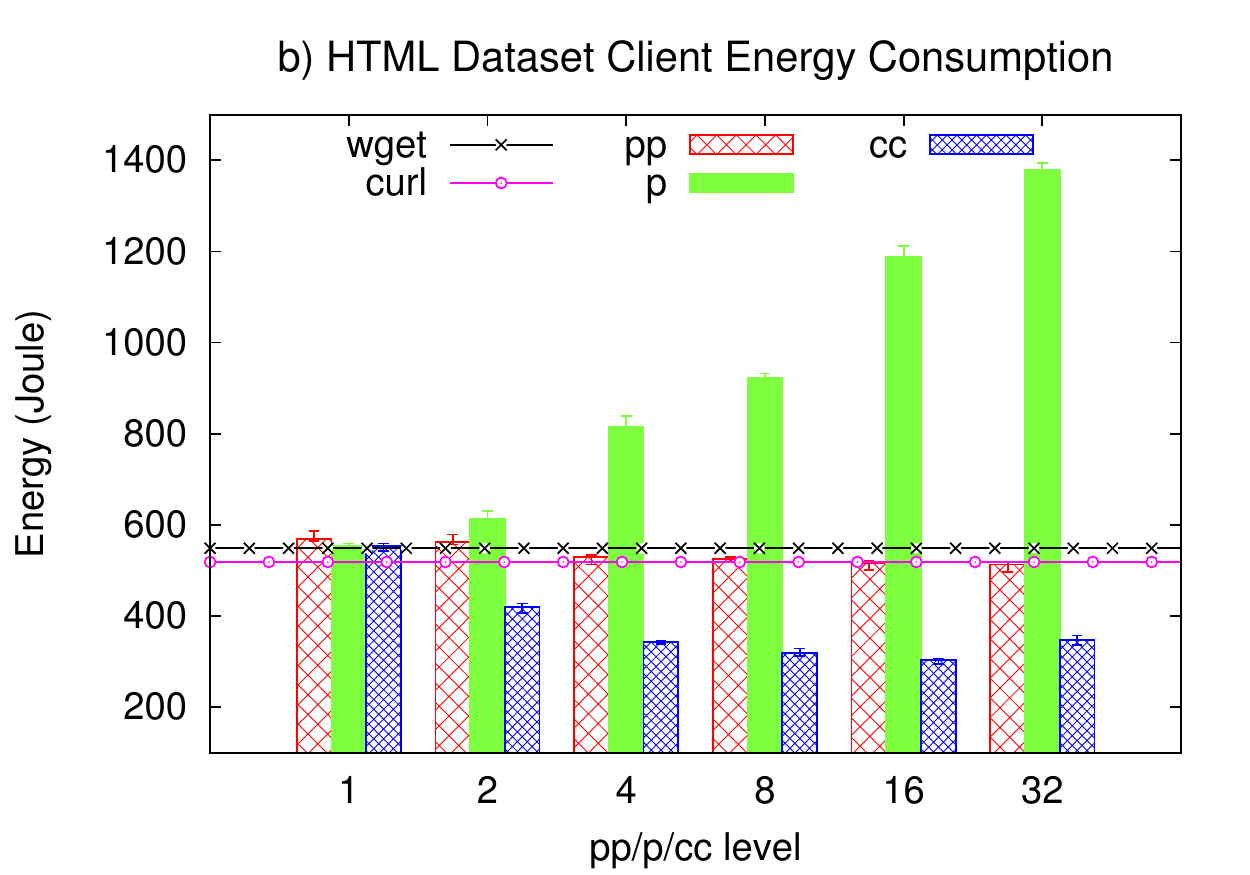}
 	\includegraphics[keepaspectratio=true,angle=0,width=55mm]{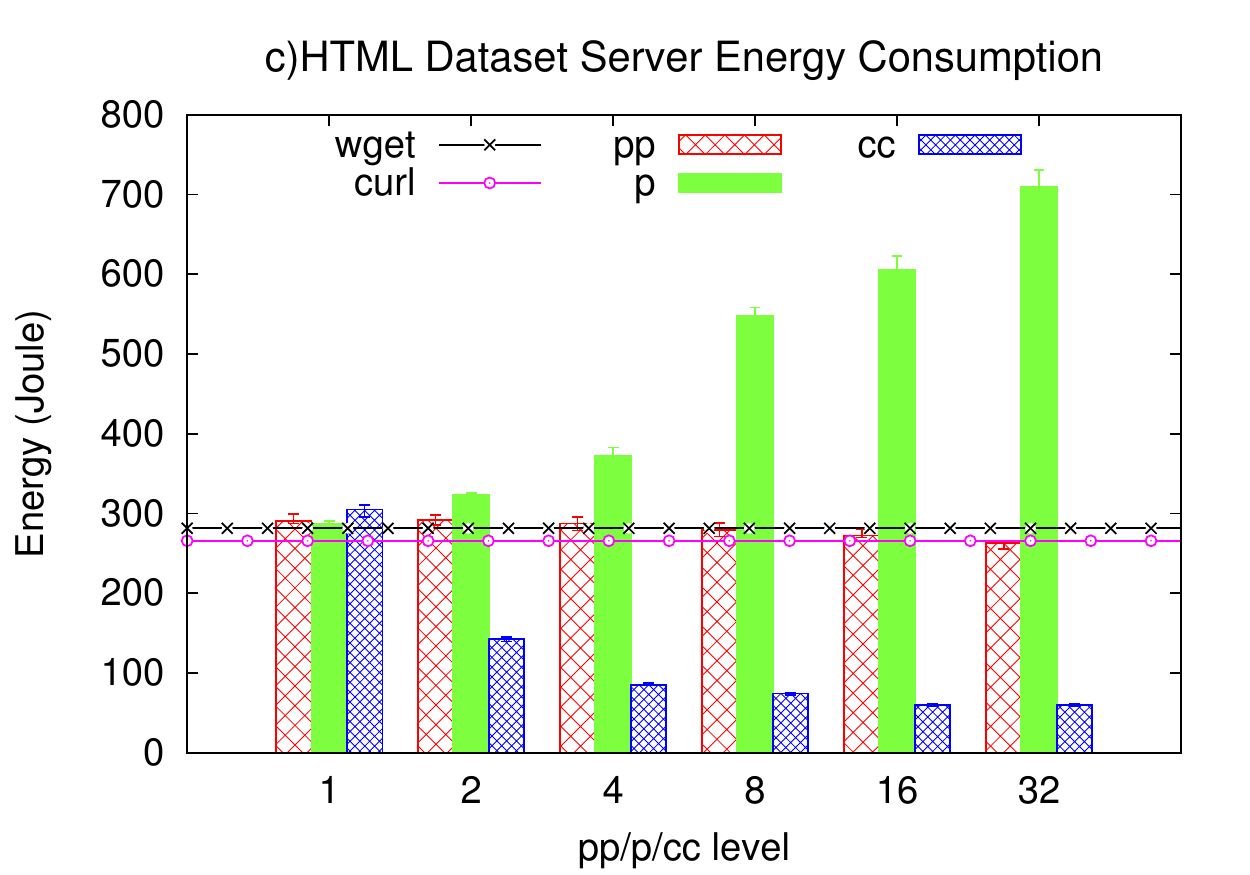}\\
	\includegraphics[keepaspectratio=true,angle=0,width=55mm]{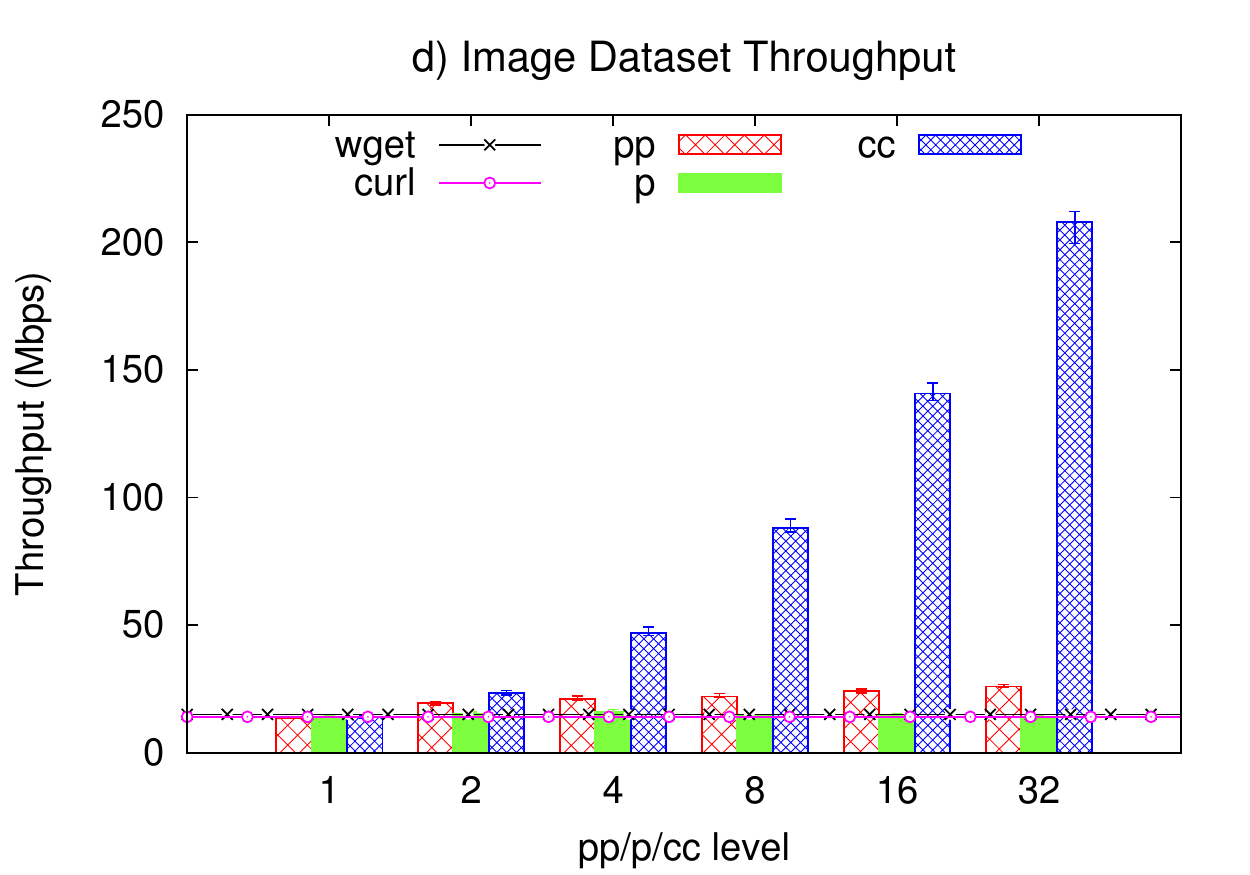}
 	\includegraphics[keepaspectratio=true,angle=0,width=55mm]{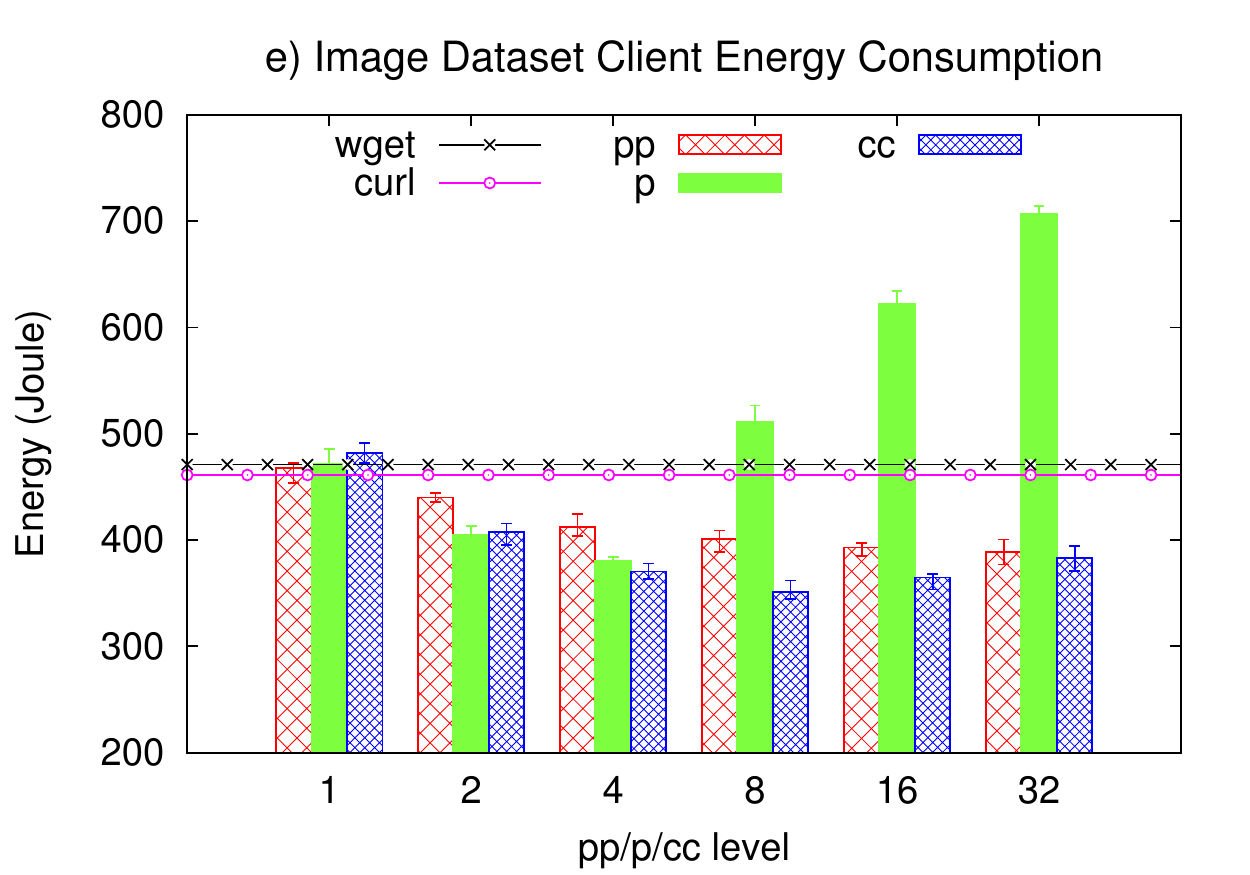}
 	\includegraphics[keepaspectratio=true,angle=0,width=55mm]{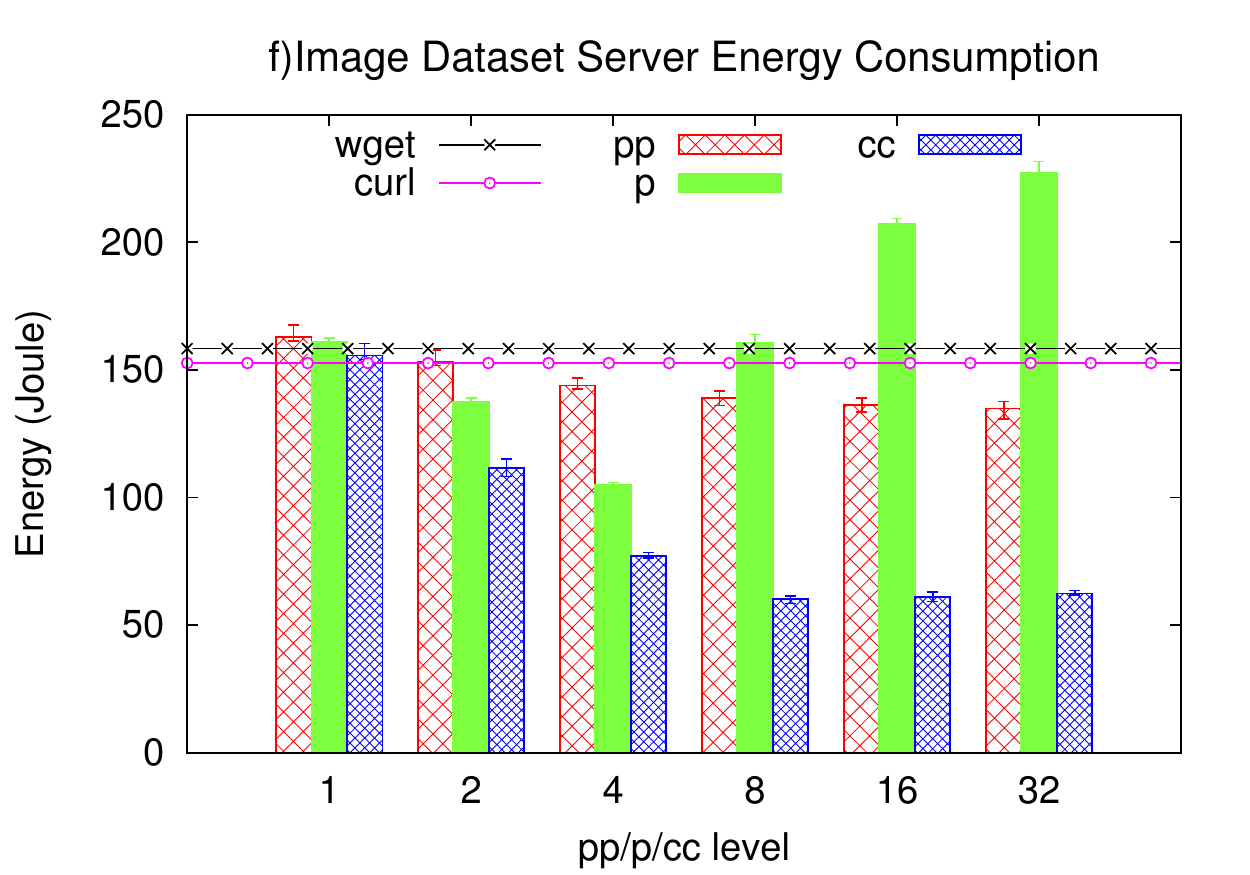}\\
 	\includegraphics[keepaspectratio=true,angle=0,width=55mm]{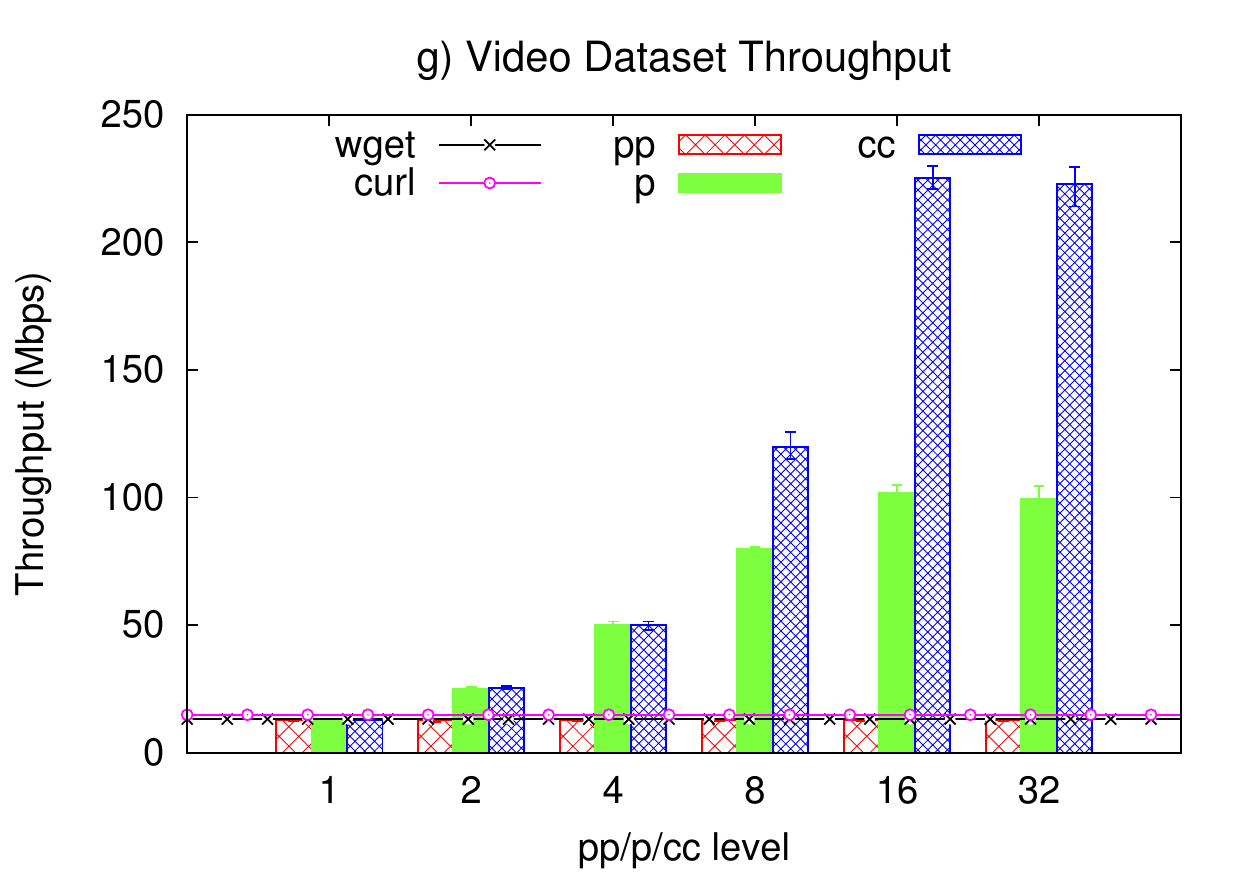}
 	\includegraphics[keepaspectratio=true,angle=0,width=55mm]{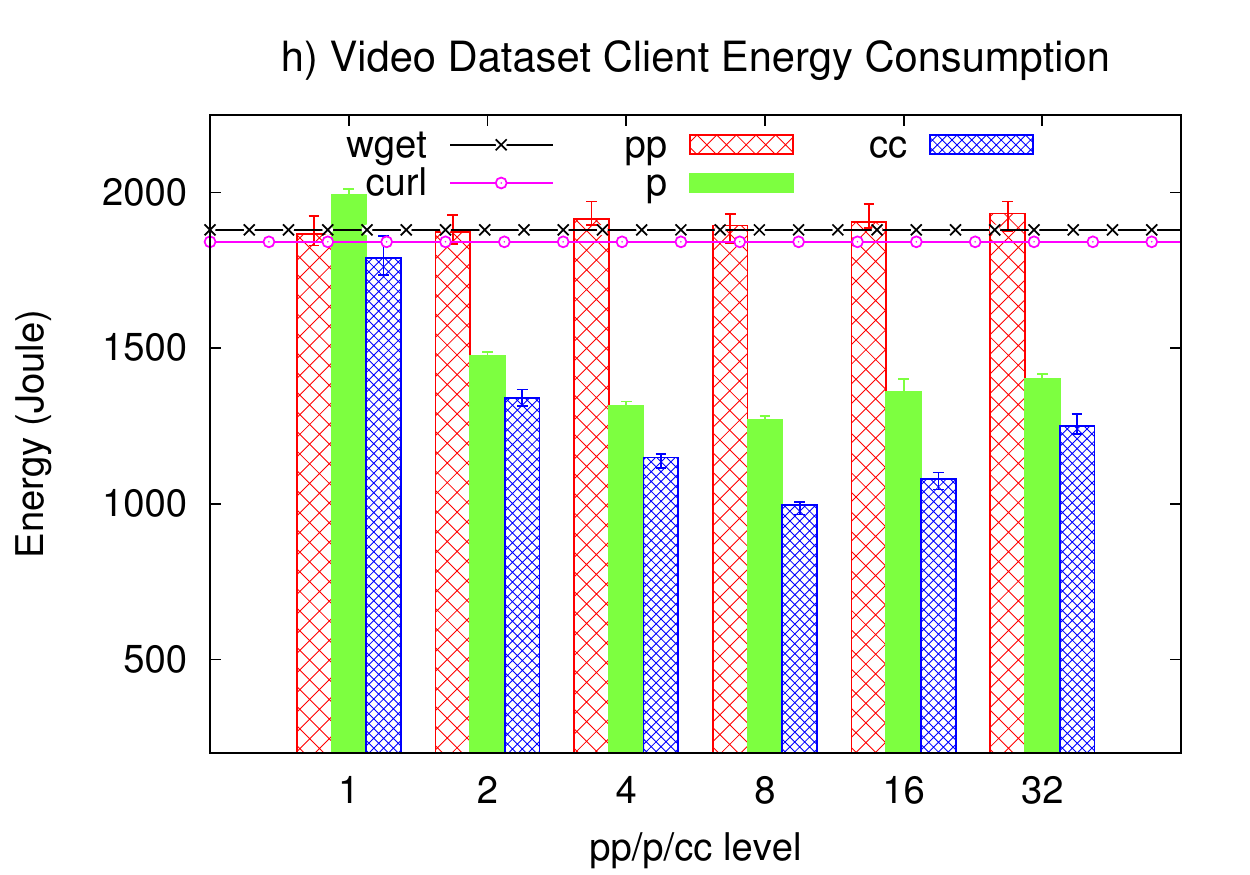}
 	\includegraphics[keepaspectratio=true,angle=0,width=55mm]{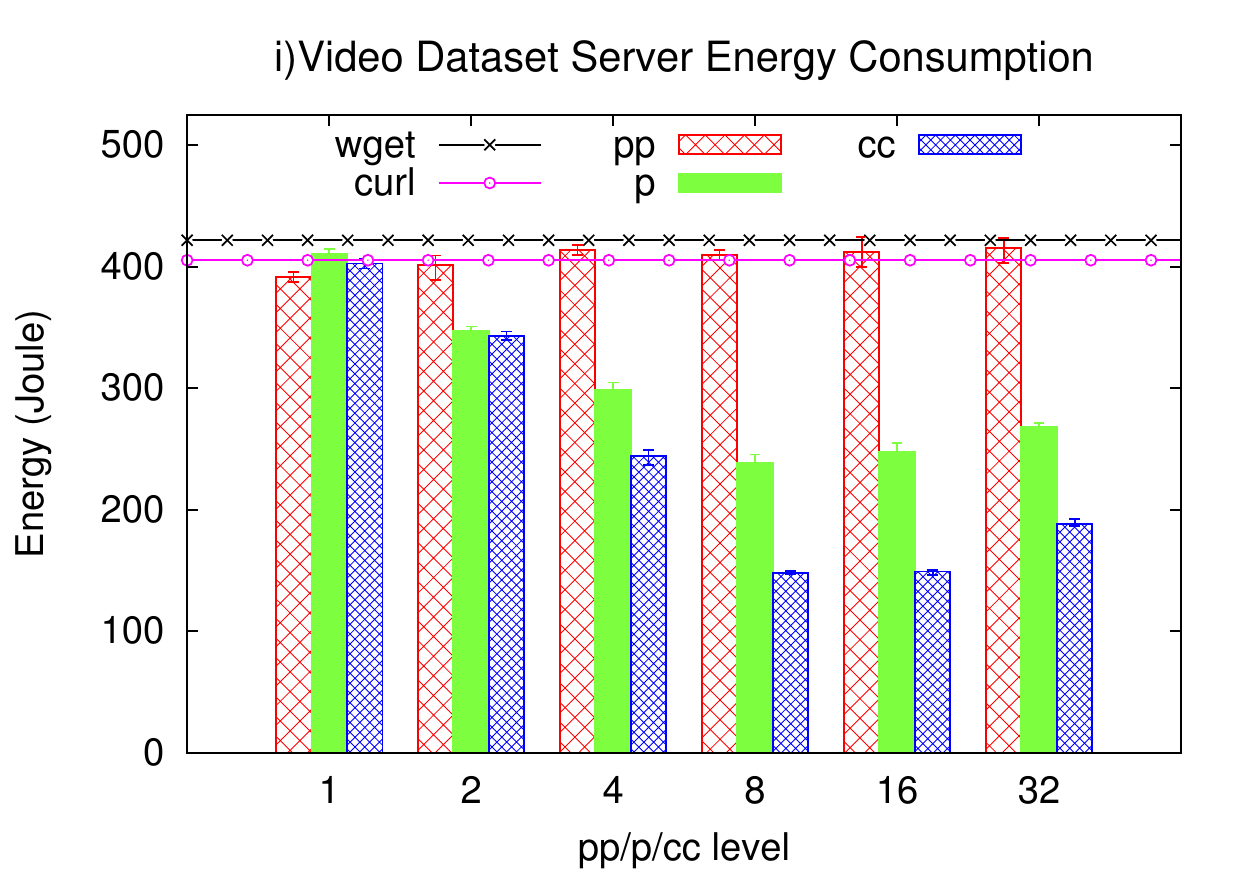}\\
\end{tabular}
\caption{Performance vs energy consumption trade-offs of HTTP transfers from a web server in Virginia to a client in Sydney.}
\label{fig:sydney-virginia}
\end{centering}
\end{figure*}

Increased levels of pipelining and concurrency improve the end-to-end HTTP data transfer throughput and decrease end-system energy consumption for the small (HTML) and medium (image) sized datasets as it is presented in Figure~\ref{fig:didclab-alamo}(a)--(f). Even though pipelining and concurrency lead to an increase in the instantaneous power consumption, decrease in the data transfer time overcomes and the overall energy consumption decreases.The pipelining increases throughput 48\% for HTML dataset and 265\% for image dataset while decreasing total energy consumption by 16\% and 20\% respectively. The concurrency is more effective than pipelining for transferring HTML and image datasets. As we increase the concurrency level, throughput jumps significantly from 9 Mbps to 102 Mbps (11.3X increase) for the HTML dataset and 16 Mbps to 174 Mbps (10.8X increase) for the image dataset. The energy consumption continues to decrease until the concurrency level reaches 16 for both datasets. After this break point, the throughput continues to increase but the energy consumption also starts to increase, since the throughput gain after this point does not overcome the overhead of doubling the number of processes at the end system. Wget and curl achieve very close transfer throughput at level 1 with our client for HTML and image datasets, however, wget consumes around 15\% and curl consumes around 10\% more energy for transferring HTML dataset and they consume same amount of energy for transferring the image dataset. When we compared wget and curl with optimal concurrency level, they consume around 40\% more energy for transferring HTML and image datasets. Increasing the level of parallelism for the same datasets decreases the transfer throughput and increases the energy consumption, since the file sizes in these datasets are too small for gaining advantage of parallelism.

For the large (video) dataset, as shown in Figure~\ref{fig:didclab-alamo}(g)--(i), pipelining has almost no effect on the throughput since the average file size is much bigger than the BDP of the network. On the other hand, both parallelism and concurrency increase the throughput and decrease the energy consumption. Parallelism increases throughput by 232\% when its value is increased from 1 to 4. Further increasing the parallelism triggers an increase in the energy consumption. As the concurrency level increases from 1 to 8, the throughput increases 6.3X and total energy consumption decreases 45\%. 

\begin{figure*}[t]
\begin{centering}
\begin{tabular}{cc}
 	\includegraphics[keepaspectratio=true,angle=0,width=55mm]{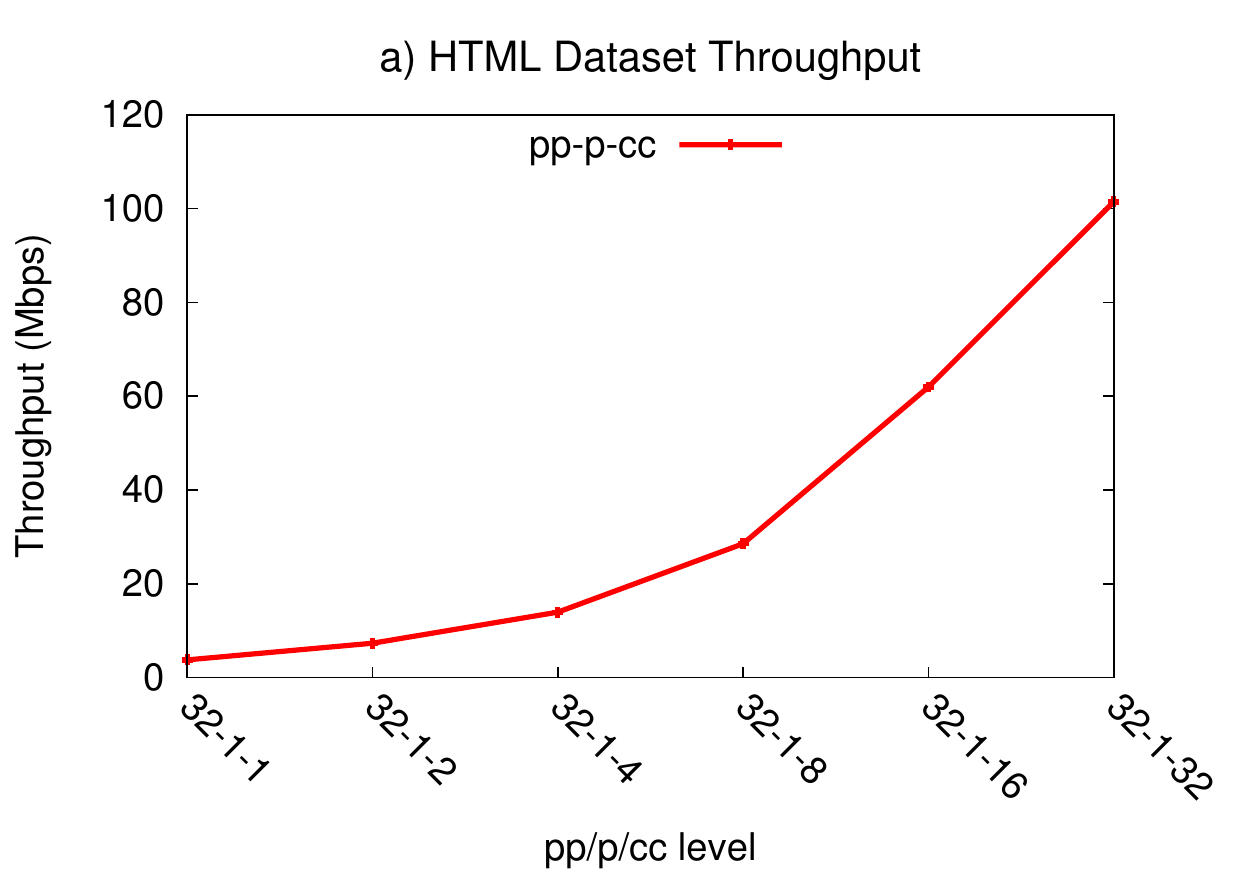}
 	\includegraphics[keepaspectratio=true,angle=0,width=55mm]{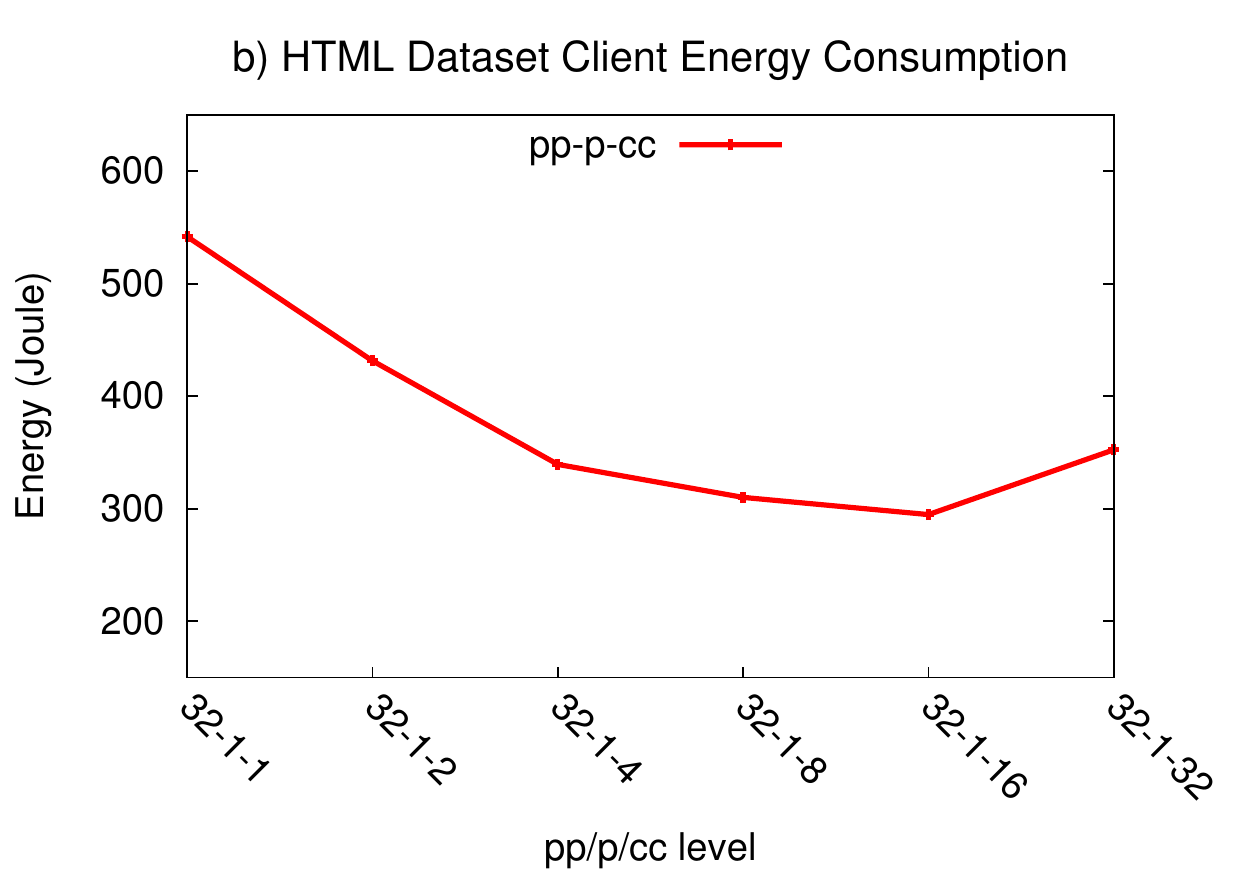}
 	\includegraphics[keepaspectratio=true,angle=0,width=55mm]{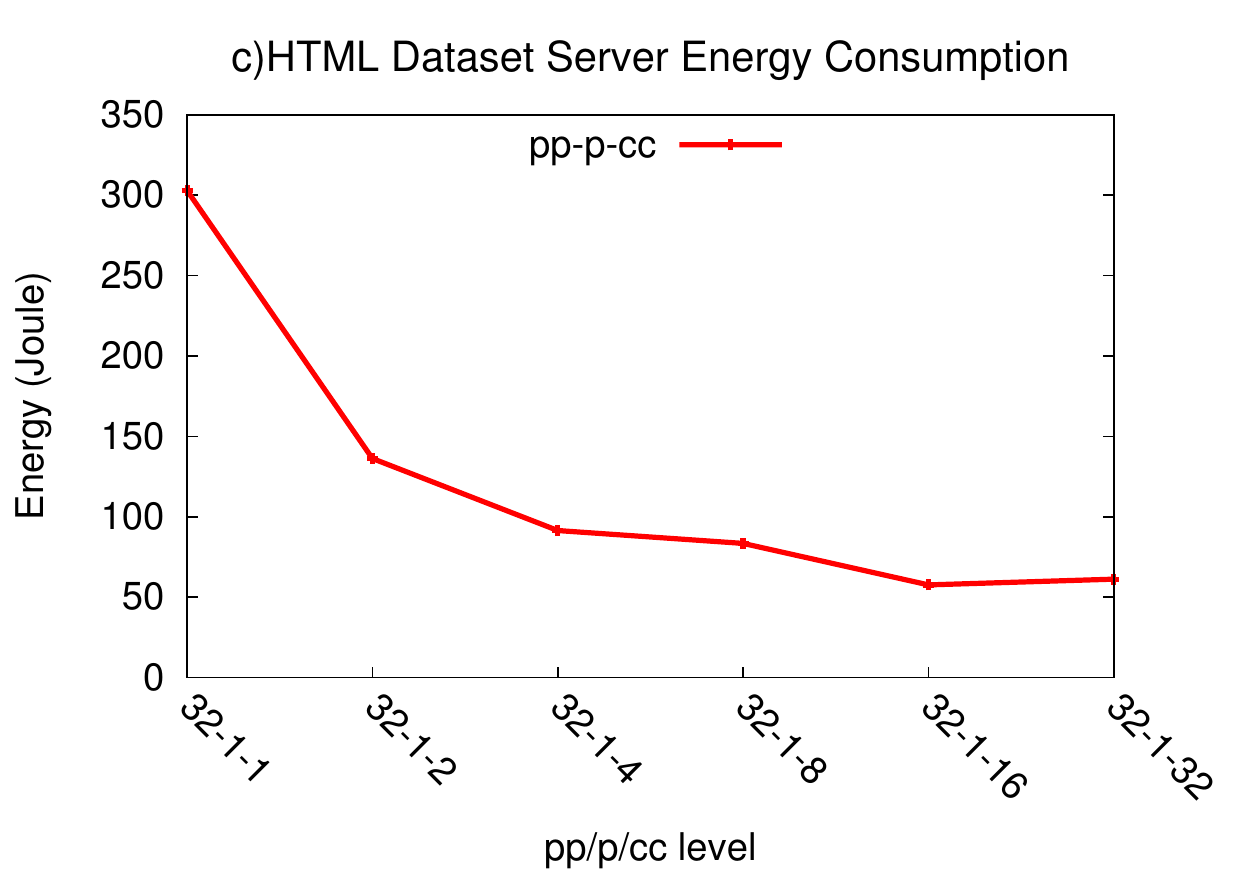}\\
	\includegraphics[keepaspectratio=true,angle=0,width=55mm]{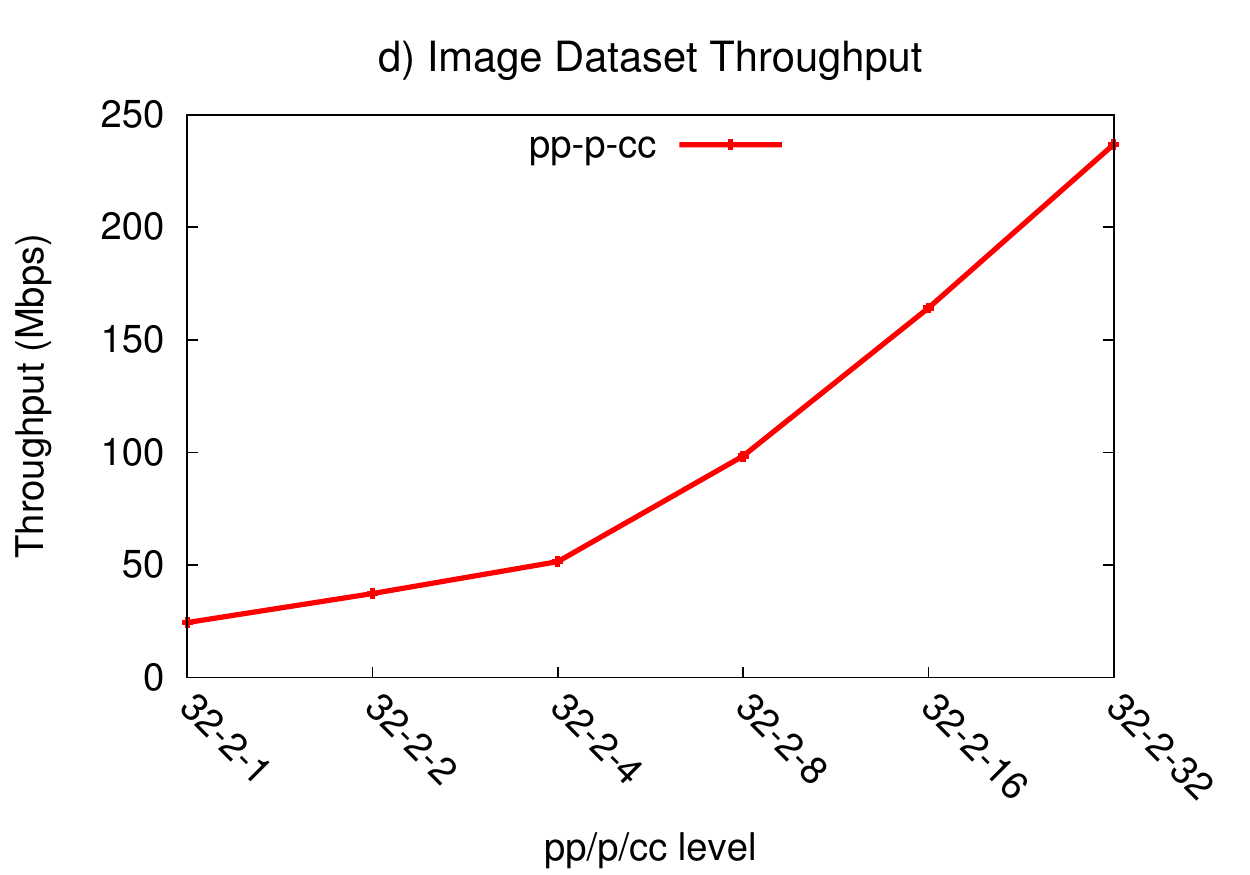}
 	\includegraphics[keepaspectratio=true,angle=0,width=55mm]{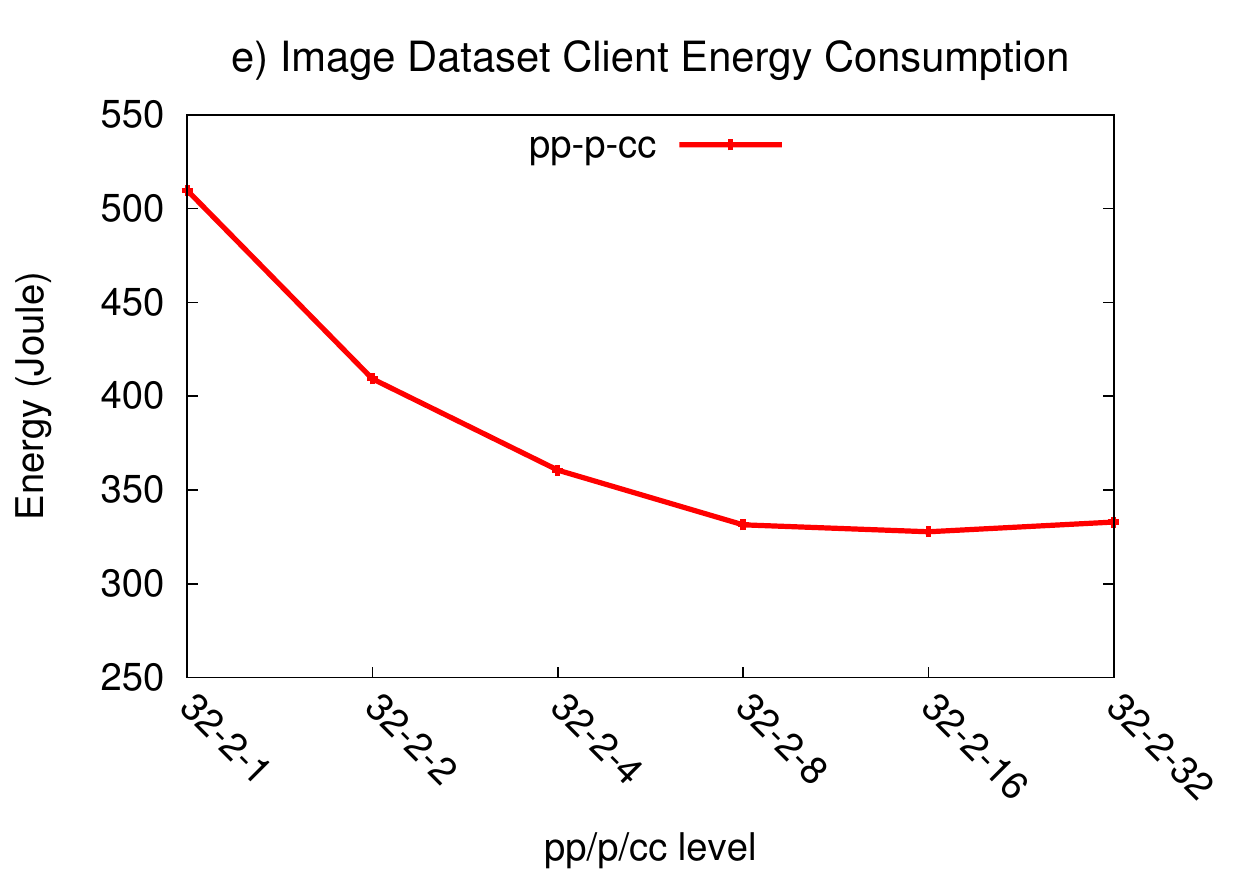}
 	\includegraphics[keepaspectratio=true,angle=0,width=55mm]{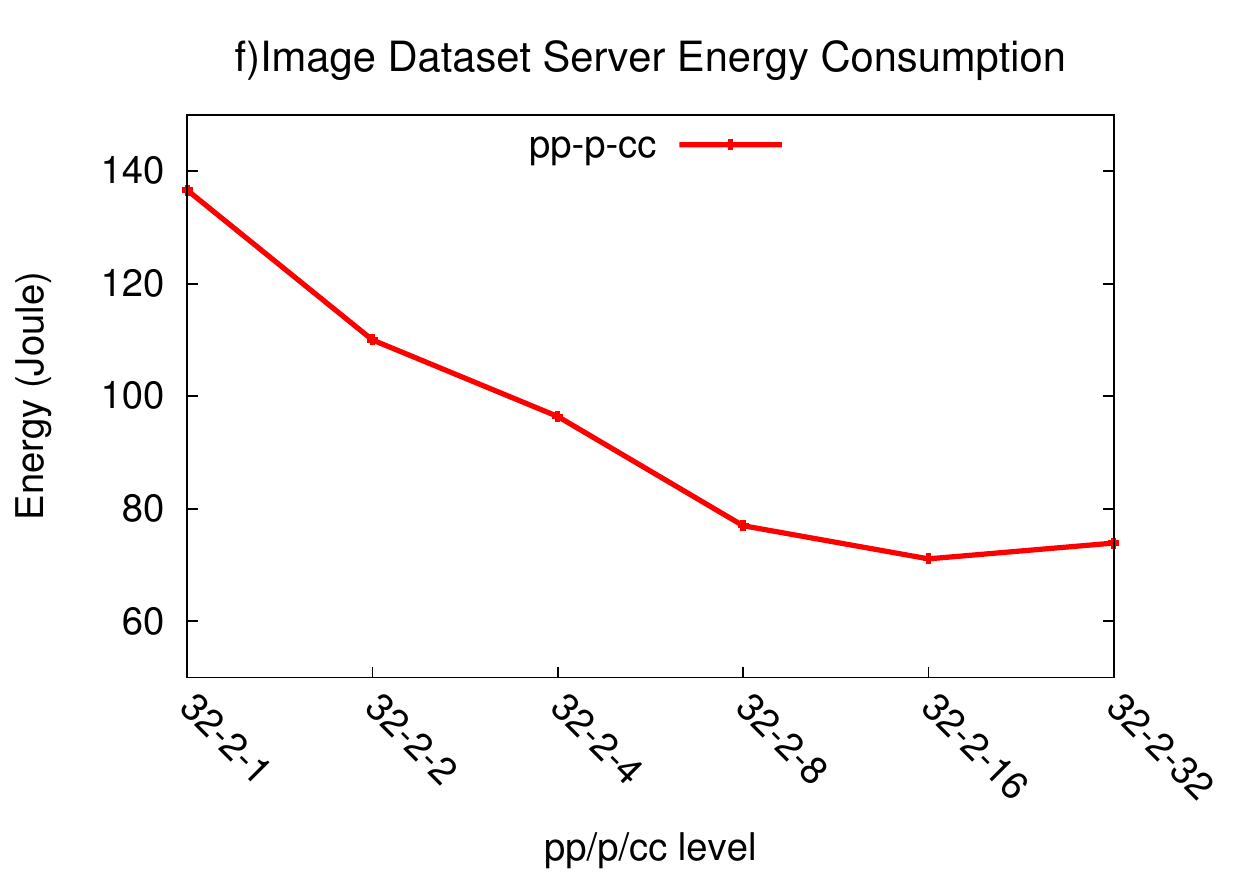}\\
 	\includegraphics[keepaspectratio=true,angle=0,width=55mm]{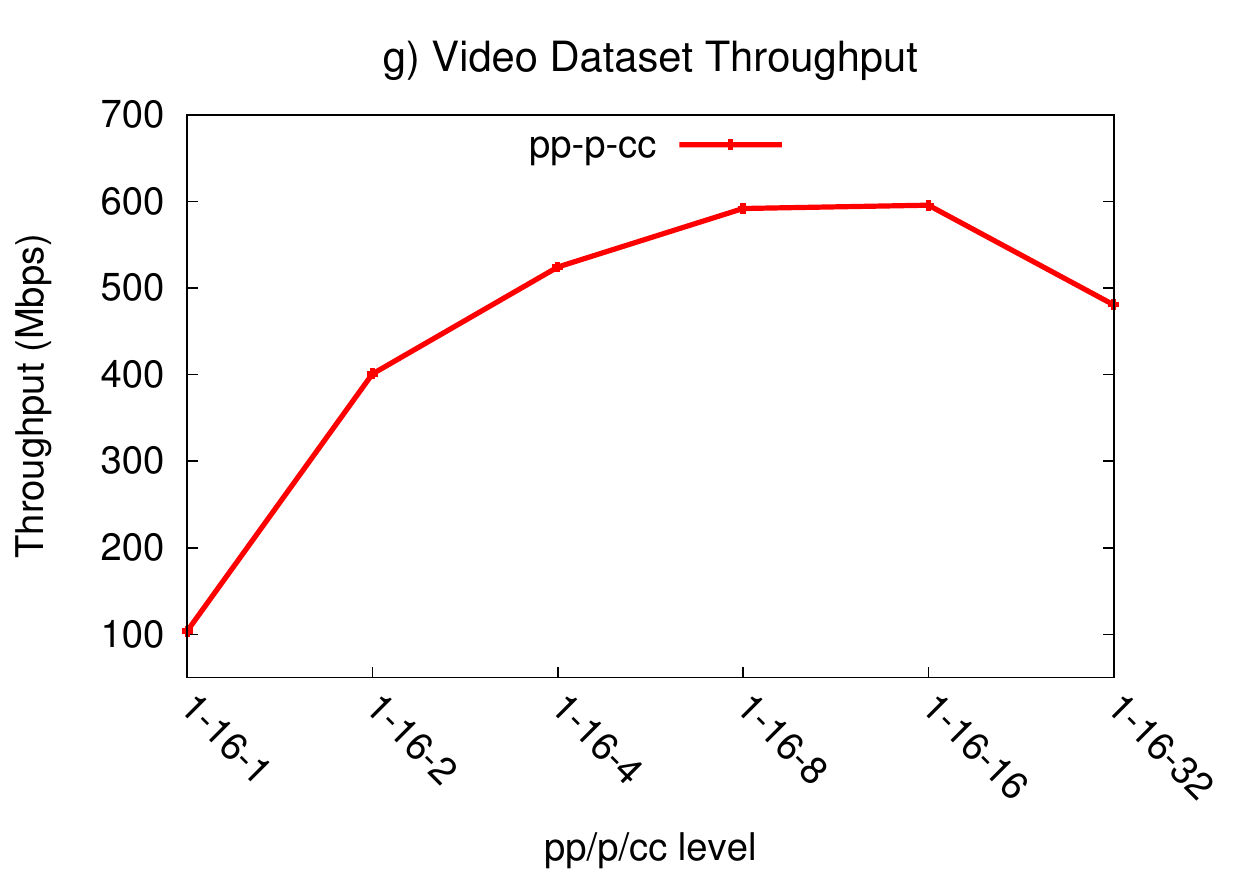}
 	\includegraphics[keepaspectratio=true,angle=0,width=55mm]{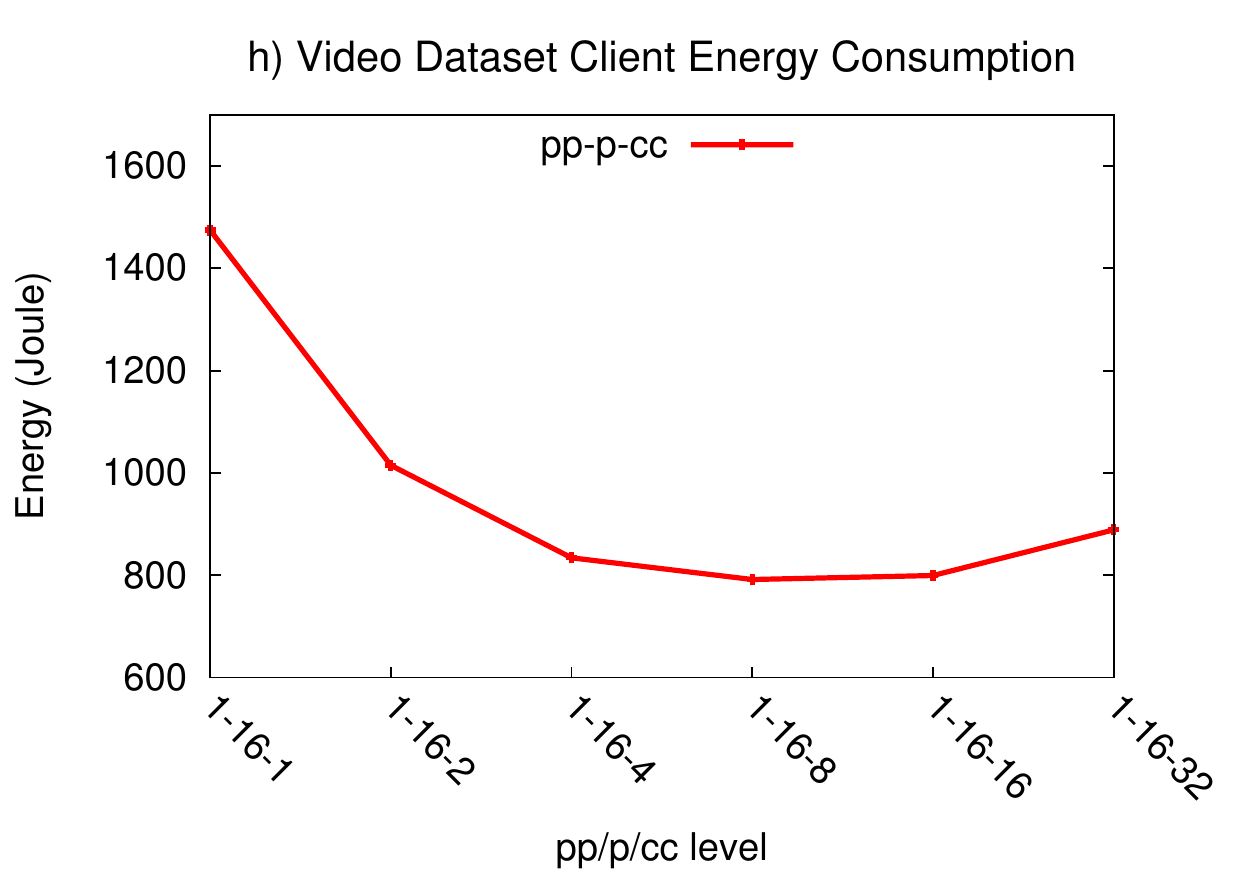}
 	\includegraphics[keepaspectratio=true,angle=0,width=55mm]{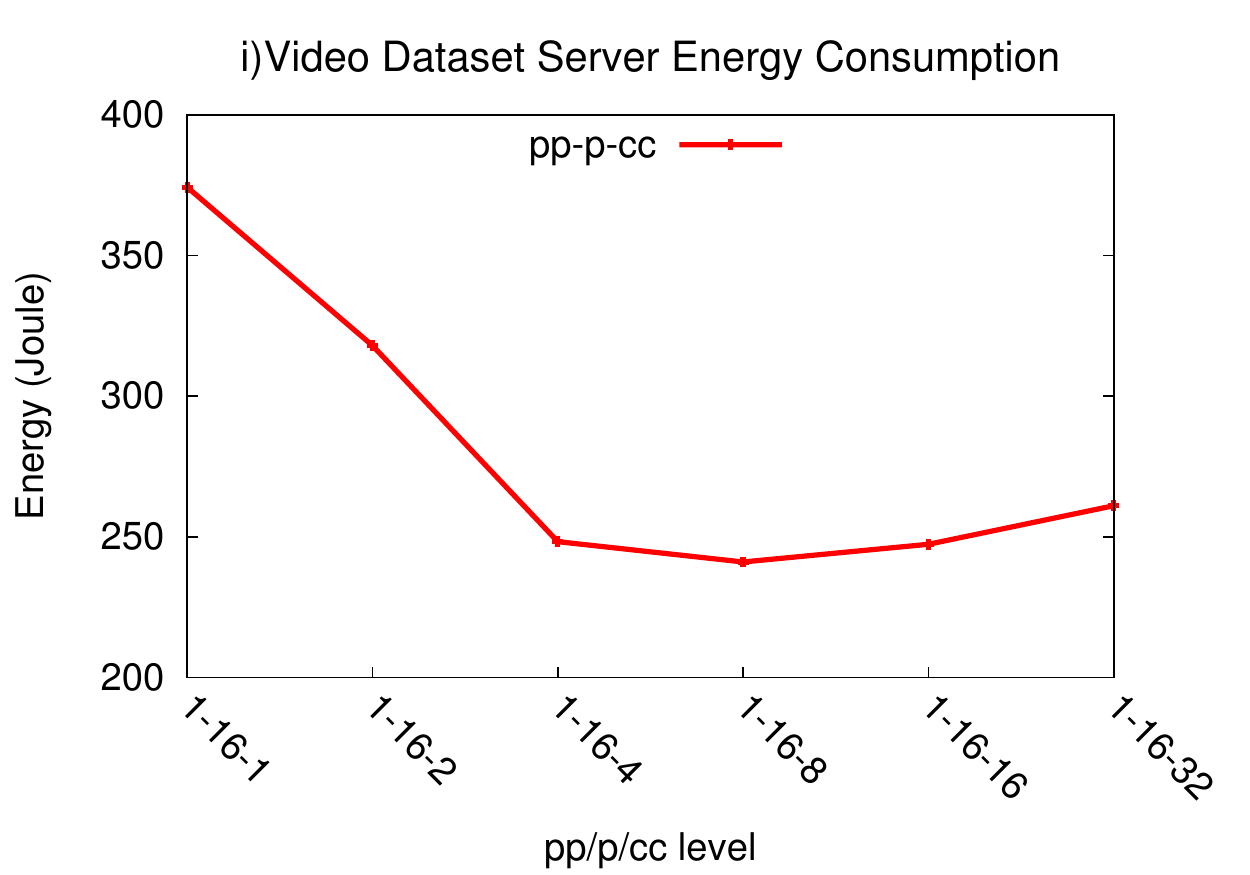}\\
\end{tabular}
\caption{Transfers with combined parameters between Sydney and Virginia.} \label{fig:sydney-virginia-combined}
\end{centering}
\vspace{-5mm}
\end{figure*}

Figure~\ref{fig:sydney-virginia} shows the effect of transfer protocol parameters for a wide area network which has a longer round trip time. We conducted these experiments between two AWS nodes where network bandwidth is 1 Gbps and RTT is 240ms. The web server is located in Virginia and web client is located in Sydney. In this setting, due to longer RTT between nodes, initial unoptimized throughput rates are around half of the previous experiment's initial transfer rates. With data transfer tuning, we can reach up to throughput ratios very close to the optimal levels. In this setting, positive effects of pipelining diminishes for the HTML and image datasets. Pipelining can increase the throughput only around 7\% and 95\% while decreasing energy consumption by 10\% and 19\% respectively from level 1 to 32, since BDP of this network is much bigger than the average file size of these datasets. 
For the video dataset, pipelining has almost no effect on throughput and energy consumption. Parallelism does not increase the throughput for the HTML and image datasets in this network condition, but it increases the energy consumption by almost 2X. However for the video dataset, it increases the throughput and decreases the total power consumption until reaching its optimal values. It increases the throughput by 6x while decreasing the power consumption by 55\% from level 1 to 8.
We can conclude that, the parallelism can improve the throughput and decrease the energy consumption of large-file dominated dataset transfers if it can be set to the optimal value. Same as the previous experiments, concurrency is the most effective parameter. As opposed to pipelining and parallelism, concurrency can increase the throughput and decrease the energy consumption considerably for all datasets. We reached the lowest energy consumption at level 16 for the HTML dataset with 45\% energy gain, at level 8 for the image dataset with 28\% energy gain, and at level 8 for the video dataset with 48\% energy gain. 

We also run transfers by using a combination of different parameters at the same time as shown in Figure~\ref{fig:sydney-virginia-combined}. The values of pipelining and parallelism are fixed to the values which returned the best throughput/energy ratio. This energy efficiency ratio quantifies how much data can be transferred at the cost of unit energy consumption. The transfer throughput significantly increases when a combination of different parameters are used compared to the case when they are used individually. 
For the HTML dataset, the combined parameter optimization increases the throughput by up to 20X while decreasing the energy consumption by up to 80\%. For the image dataset, the throughput is increased by up to 10X, and the energy consumption is reduced by up to 50\%. 
Finally, for the video dataset, the throughput is increased by up to 6X, and the energy consumption is reduced by up to 40\%. 
These results show that using a good combination for these three transfer parameters increases the end-to-end HTTP data transfer throughput drastically while reducing the total energy consumption at the sender and receiver nodes significantly.

\begin{figure}[t]
\centerline{%
\includegraphics[width=0.25\textwidth,height=0.14\textheight]{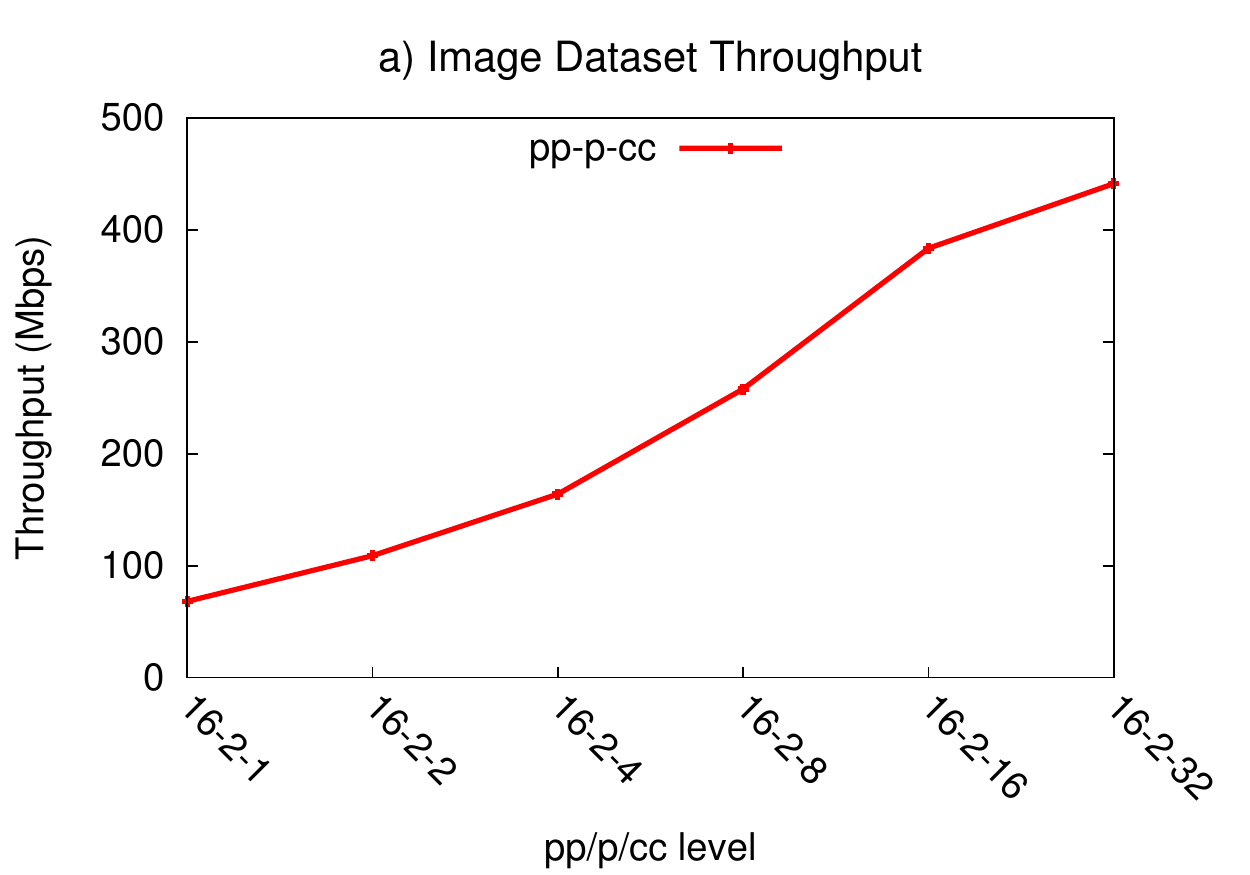}%
\includegraphics[width=0.25\textwidth,height=0.14\textheight]{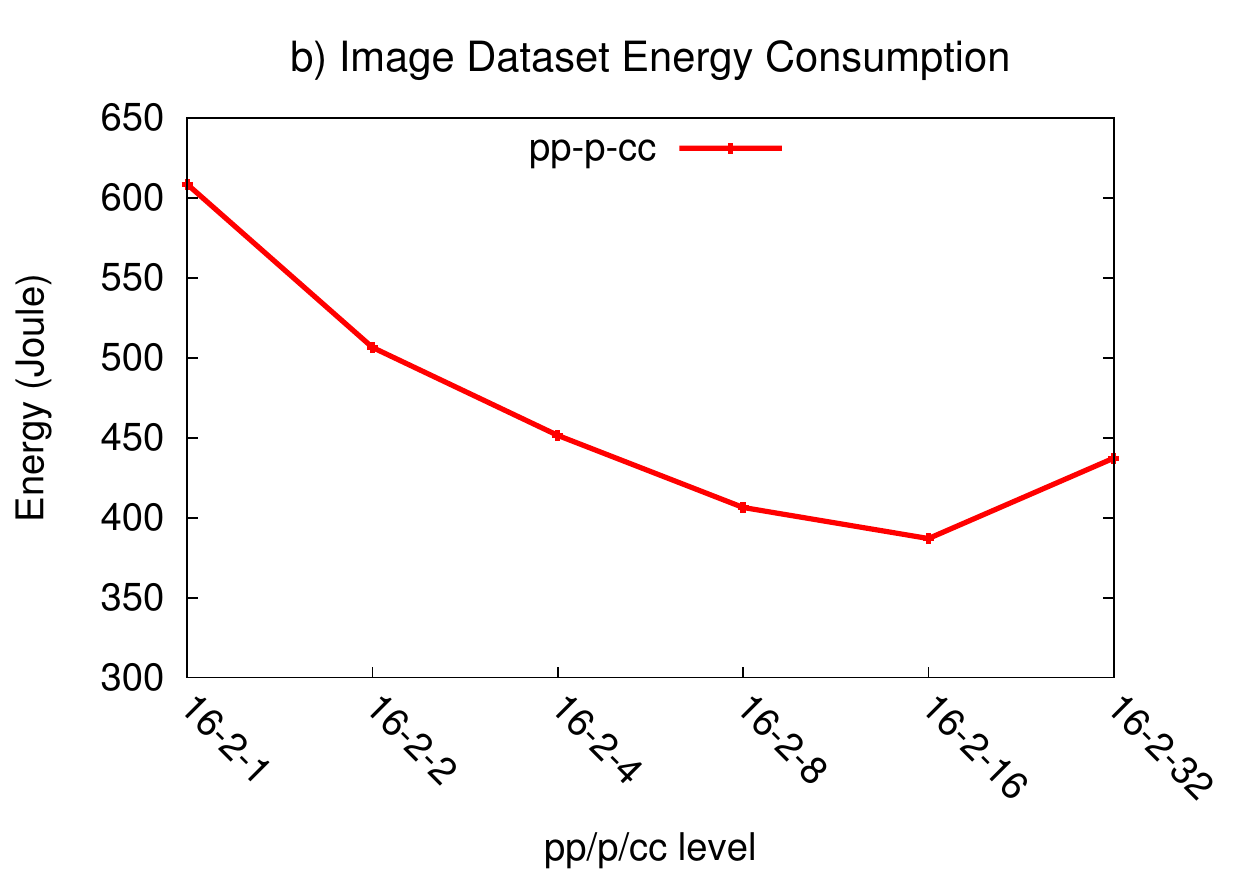}%
}%
\centerline{%
\includegraphics[width=0.5\textwidth,height=0.12\textheight] {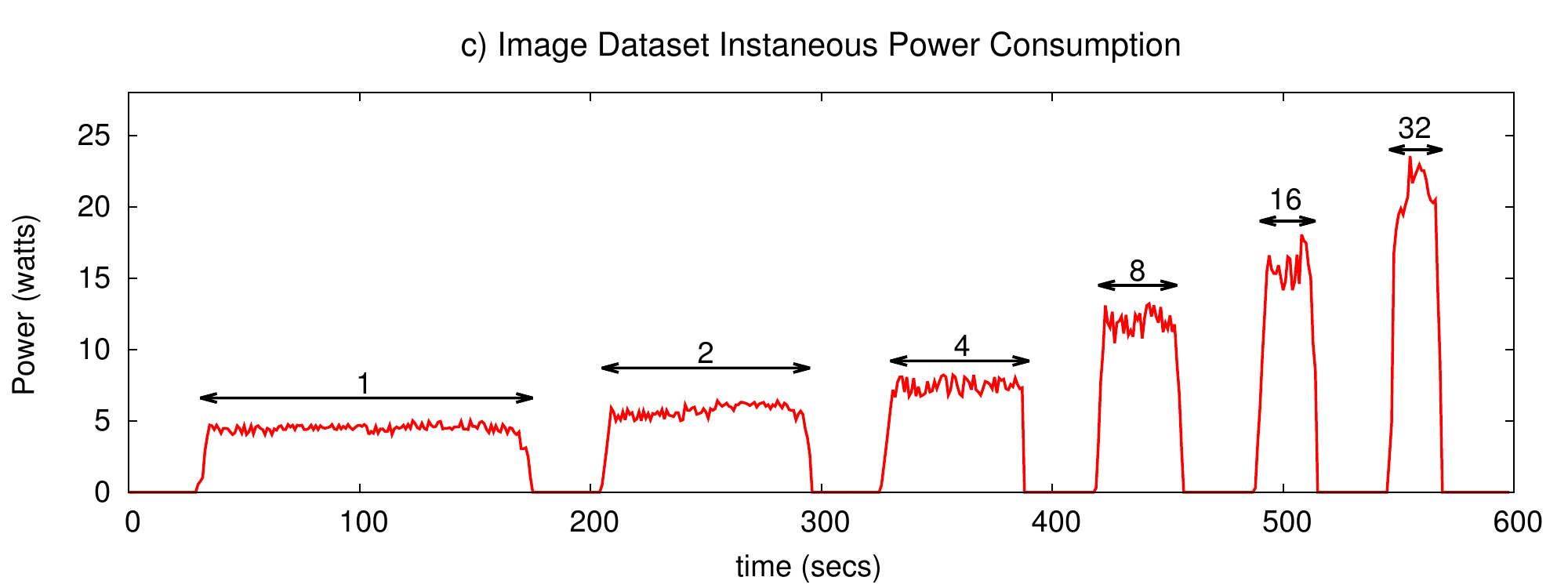}%
}%
\caption{Effect of combined parameters on (a) end-to-end throughput and (b) total energy consumption (for the image dataset transfers between Oregon and Virginia AWS EC2 nodes); (c) concurrency level versus instantaneous power consumption.}
\label{fig:saveEnergy}
\end{figure}

In Figure~\ref{fig:saveEnergy}, we can see how the instantaneous power consumption changes during the image dataset transfer between Oregon and Virginia with combined parameters. We included only the image dataset transfer results in this case due to space limitations and also the results for other datasets were quite similar. In this experiment, the pipelining level is fixed to 16, and the parallelism level is fixed to 2 (since these are the best values for this specific setting), whereas the concurrency level is changed between 1 and 32. When we increase the level of concurrency from 1 to 32, the average instantaneous power consumption increases from 4.2 watts to 19.5 watts, whereas the transfer time decreases from 145 seconds to 23 seconds. Thus overall energy consumption substantially decreases from 608 joule to 437 joules. This graph helps us to understand the nice balance between the increased instantaneous  power consumption and the decreased transfer time, since both affect the total energy consumption of the system.
As long as the energy gain due to the decreased transfer time is more than the loss due to the increased instantaneous power consumption, then we
save energy at this system while increasing the throughput. But this is not always the case. In some transfers, we observe that although
the throughput continues to increase, the total energy consumption would not continue to decrease, instead come to a balance and start
increasing again. This is also what we observe in Figure~\ref{fig:saveEnergy}(b), when the concurrency level is increased from 16 to 32. 

{\bf Lessons Learned.} The lessons learned from these experiments that should be harnessed while developing energy-aware data transfer algorithms can be summarized as follows:
{\em (1)} Concurrency is the most effective and helpful parameter in terms of increasing the throughput and decreasing the total energy consumption, but the optimal concurrency value can be different depending on the dataset characteristics, end-system resources, and network conditions. Increasing the concurrency level does not always lead to energy savings and throughput gain, and estimating that fine point is a challenging task.
{\em (2)} Pipelining leads to energy savings depending on the file size and BDP. If the average file size is less than BDP, pipelining increases the throughput and leads  to energy savings. For larger file sizes, pipelining does not effect the throughput or energy consumption much.
{\em (3)} High levels of parallelism can cause energy lost and throughput decline for small files but it is very effective for transferring large files such as the video dataset in our experiments.

\section{SLA-Based Transfer Algorithms}

Service Level Agreements (SLA) have become more popular with the rapidly advancing and growing pay-as-you-go and low cost cloud computing services. As the usage of such systems increases, the guaranteed reliability and quality of the provided services become more important, which requires mutually agreed Service Level Agreements (SLA). Hence, we devised various HTTP data transfer algorithms based on different SLA requirements considering the lessons learned from our previous experiments.

\subsection{SLA: Minimum Energy Consumption}
The aim of this algorithm is to minimize the energy consumption of data transfers by tuning the analyzed transfer parameters. The SLA: Minimum Energy (MinE) algorithm transfers files with minimum energy consumption without considering any performance criteria. In other words, we offer an option for the users who do not need to finish the transfers in the time pressure.  We believe this approach will be helpful for background transfers and update procedures which do not require any time limitations.

\begin{algorithm}[t]
\caption{--- SLA: Minimum Energy Consumption} \label{alg:MinEnergy}
\algsetup{
linenosize= \scriptsize,
linenodelimiter=.
}
\begin{algorithmic} 
	\REQUIRE $Channel Count$
	\STATE $Group Files(filelist,BDP)$
	\FOR { $each \ group \ small::large$}
		\STATE$pipelining = \left\lceil\frac{BDP}{avgFileSize}\right\rceil$ \label{line:pipelining}
		\STATE$parallelism=Min(\left\lceil\frac{BDP}{bufSize}\right\rceil,\left\lceil \frac{avgFileSize}{bufSize}\right\rceil)$ \label{line:parallelism}
		\STATE $concurrency= Min(\left\lceil\frac{BDP}{avgFileSize}\right\rceil,\left\lceil\frac{ChannelCount+1}{2}\right\rceil)$			                        \label{line:concurrency}
		\STATE$Channel Count = Channel Count - concurrency$		
		
	\ENDFOR
	\STATE $startTransfer(groups)$ \label{minenergy:transfer}
\end{algorithmic}
\end{algorithm}

Instead of using the same parameter combination for the whole data set, which can increase the power consumption unnecessarily, we initially group files into three subgroups; {\em Small}, {\em Medium} and {\em Large} based on the file sizes and the Bandwidth-Delay-Product (BDP). Then, we calculate the best possible parameter combinations for each subgroup in which BDP, average file size, and TCP buffer size are taken into consideration. Pipelining and concurrency are the most effective parameters for small file transfers, so it is especially important to choose the best pipelining and concurrency values for such transfers. Pipelining value is calculated by dividing BDP to the average file size of the group which returns large values for {\em Small} files. By setting pipelining to relatively high values, we are transferring multiple data packets back-to-back, which in turn prevents idleness of the network and system resources, and decreases the energy consumption. As the average file size of the subgroups increases, pipelining value is set to smaller values since it does not further improve the data transfer throughput. It could even cause a performance degradation and redundant power consumption by poorly utilizing the network and system resources.
Moreover, we assigned most of the available data channels to the {\em Small} subgroup which multiplies the impact of energy saving when combined with pipelining. Besides pipelining, minimum energy algorithm also tries to keep the concurrency level of {\em Large} files at minimum since using more concurrent channels for large files causes more power consumption. For the parallelism level, we again consider TCP buffer size, BDP, and average file size. The equation will return small values for {\em Small} files which will avoid creating unnecessarily high number of threads and thus prevents redundant power consumption. The parallelism level for {\em Medium} and {\em Large} subgroups will be high if the system buffer size is not large enough to fill the channel pipe.

\subsection{SLA: Maximum Throughput}

In the SLA: Maximum Throughput (MaxThr) algorithm, the focus is mainly maximizing the overall throughput without considering power consumption. As we mentioned in Section 3, even after choosing the best parameter combination for each file, the throughput obtained during the transfer of the small files is significantly lower compared to the large files due to the high overhead of reading too many files from disk and under utilization of the network pipe. Hence, keeping the balance between large and small files in the transfer of heterogeneous datasets is important for achieving the maximum throughput.

\begin{algorithm}[t]
\caption{--- SLA: Maximum Throughput} \label{alg:MaxThr}
\begin{algorithmic} 
	\REQUIRE $Channel Count$
	\STATE $GroupFiles(filelist,BDP)$
	\STATE $calculateOptimalParameters()$
	\FOR { $i=0$ to $ChannelCount$}
		\IF{$i\%=0$}
			\STATE{$large \ concurrency ++$}
		\ELSIF{$i\%=1$}
			\STATE{$medium \  concurrency ++$}
		\ELSE
			\STATE{$small \ concurrency ++$}
		\ENDIF
	\ENDFOR
	\STATE $startTransfer(groups)$ 
\end{algorithmic}
\end{algorithm}

Similar to MinE, MaxThr algorithm groups files into three subgroups according to the file size and calculates the best possible parameter values for pipelining and parallelism using BDP, TCP buffer size and average file size of the subgroup. The algorithm distributes data channels among subgroups using round-robin algorithm in the order of Large-Medium-Small. After channel distribution is completed, algorithm transfers subgroups concurrently using the calculated concurrency level for each subgroup. The algorithm continues to check the channel distribution until completion of all file transfers. When the transfer of all files in a subgroup is completed, the channels of the subgroup are scheduled for another subgroup based on same round-robin order.   

\subsection{SLA: Energy Efficiency}
Energy efficiency is defined as using less energy to provide the same level of service. When we refine this general definition to the data transfer service, energy efficient data transfer means transferring data with less energy providing the same or higher throughput rates compared to the regular transfers. Hence, the main purpose of SLA: Energy Efficiency algorithm (EE) is finding the best possible parameter configuration to reach maximum throughput with minimum energy consumption. The algorithm does not only focus on minimizing the energy consumption or maximizing the throughput, rather it aims to find high performance and low power consumption concurrency levels within defined transfer channel range. While the minimum value for the concurrent number of the transfer channels is 1 for all environments, the maximum value might be different due to variability of end-system resource capacities and fairness concerns. Thus, we let the user to be able to decide the maximum acceptable number of concurrent channels for a data transfer. We have chosen concurrency to leverage the throughput and energy consumption, since in our previous experiments we have observed that concurrency is the most influential transfer parameter for all file sizes in most of the experimented settings.

After estimating the optimal pipelining and parallelism values with previous equations, we also calculated proportions for each subgroup based on total size and the number of files. Given the maximum allowed channel count, proportions are used to determine the number of channels to be allocated for each subgroup. For example, if we have a dataset dominated by small files, then assigning equal number of channels to subgroups would cause sub-optimal transfer throughput since large files will be transferred faster than smaller files and the average transfer throughput will be subjected to small subgroup's throughput. To solve this problem, we initially calculate the proportions of the subgroups and then allocate channels to the subgroups accordingly. 

The EE algorithm starts to transfer files with one active channel, and then increases the channel count 
until reaching the user-defined maximum channel count. Instead of evaluating the performance of all concurrency levels in the search space, EE halves the search space by incrementing the concurrency level by four each time. Each concurrency level is run for five-second time intervals and then the power consumption and throughput of each interval are calculated. Once EE examines \ratiotag~ratio of all concurrency levels in the search space, it picks the concurrency level with maximum \ratiotag~ratio to transfer the rest of the files.  

\begin{algorithm}[t]
\caption{--- SLA: Energy-Efficiency} \label{alg:EE}
\begin{algorithmic} 
	\REQUIRE $Maximum Channel Count$
	\STATE $GroupFiles(filelist,BDP)$
	\STATE $calculateOptimalParameters()$
	\STATE $calculateProportions$
	\COMMENT{Calculate proportional weight of each subgroup}
	
	\STATE $Channel=1$
	\WHILE{$Channel<=MaximumChannelCount$}
		\STATE $transfer(Channel)$		
		\STATE $energyConsumption=calculate(SystemMetrics)$
		\STATE $throughput=calculateThroughput()$
		\COMMENT{Calculate energy and throughput for 5 secs period.}
		\STATE $energyEfficiency[i]=\frac{throughput}{energyConsumption}$
		\STATE $Channel+=4 $
	\ENDWHILE
	\STATE $transfer(EnergyEfficientConcurrencyLevel)$ \label{ee:transfer}
\end{algorithmic}
\end{algorithm}

\subsection{SLA: Flexible Throughput by Energy Efficient Approach}

In addition to the previous algorithms, the users may want to transfer data within flexible time requirements by an energy efficient approach. Hence, we devised the SLA: Flexible Throughput (FlexibleThr) algorithm which lets the users to define their throughput requirements as a percentage of the maximum achievable throughput in a given transfer environment. The FlexibleThr algorithm takes the desired throughput value as input and aims to achieve it with minimal energy consumption by tuning the transfer parameters. While keeping the quality of service (transfer throughput in this case) at the desired level, FlexibleThr algorithm uses a technique similar to the MinE algorithm to keep the power consumption at  the minimum possible level. 

\begin{algorithm}[t]
\caption{--- SLA: Flexible Throughput by Energy Efficient Approach} \label{alg:Flexible}
\begin{algorithmic}
	\REQUIRE $Target Throughput$
	\STATE $GroupFiles(filelist,BDP)$
	\STATE $calculateOptimalParameters()$
	\STATE $startTransfer()$	
	\STATE $Throughput=calculateThroughput()$
	\COMMENT{Calculate throughput for every 5 secs period.}
	\IF{$Throughput<=targetThroughput$}
		\STATE $concurrency=targetThroughput/actThroughput$	
	\ENDIF
	\WHILE{$Throughput<=targetThroughput$}
		\STATE $concurrency++$
	\ENDWHILE	
\end{algorithmic}
\end{algorithm}

The FlexibleThr algorithm initiates the transfer with one channel, and if the throughput is less than the target throughput, it increases the required concurrency level by comparing the current and target throughput values. FlexibleThr continues to check the throughput every five seconds, and if the throughput is still less than the target throughput at the estimated concurrency level, then additive increase is applied to reach the target throughput until it reaches or exceeds the target. When assigning the transfer channels to the subgroups, FlexibleThr gives priority to the small files similar to the MinE algorithm due to the energy consumption concerns.

\begin{figure*}[t]
\begin{centering}
\begin{tabular}{cc}
\includegraphics[keepaspectratio=true,angle=0,width=55mm] {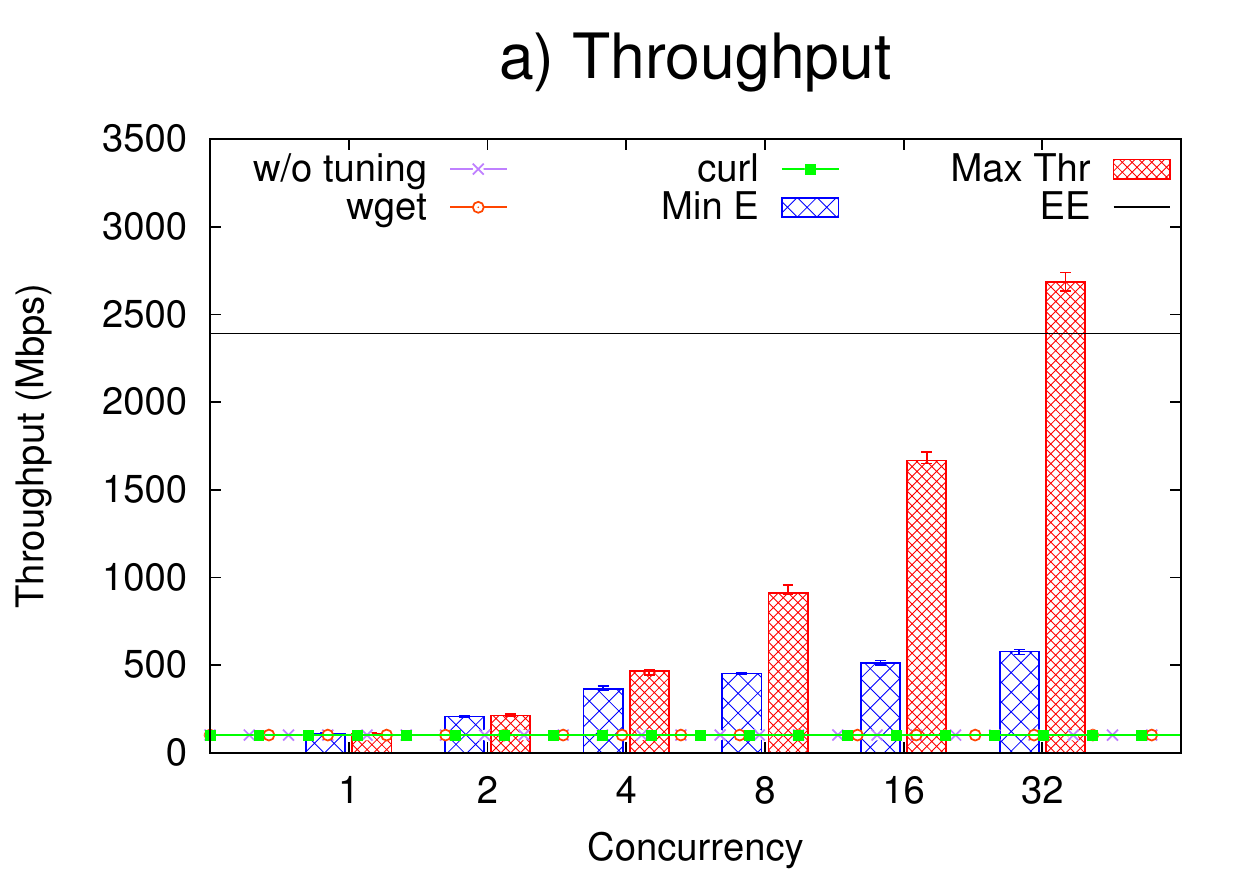}
\includegraphics[keepaspectratio=true,angle=0,width=55mm] {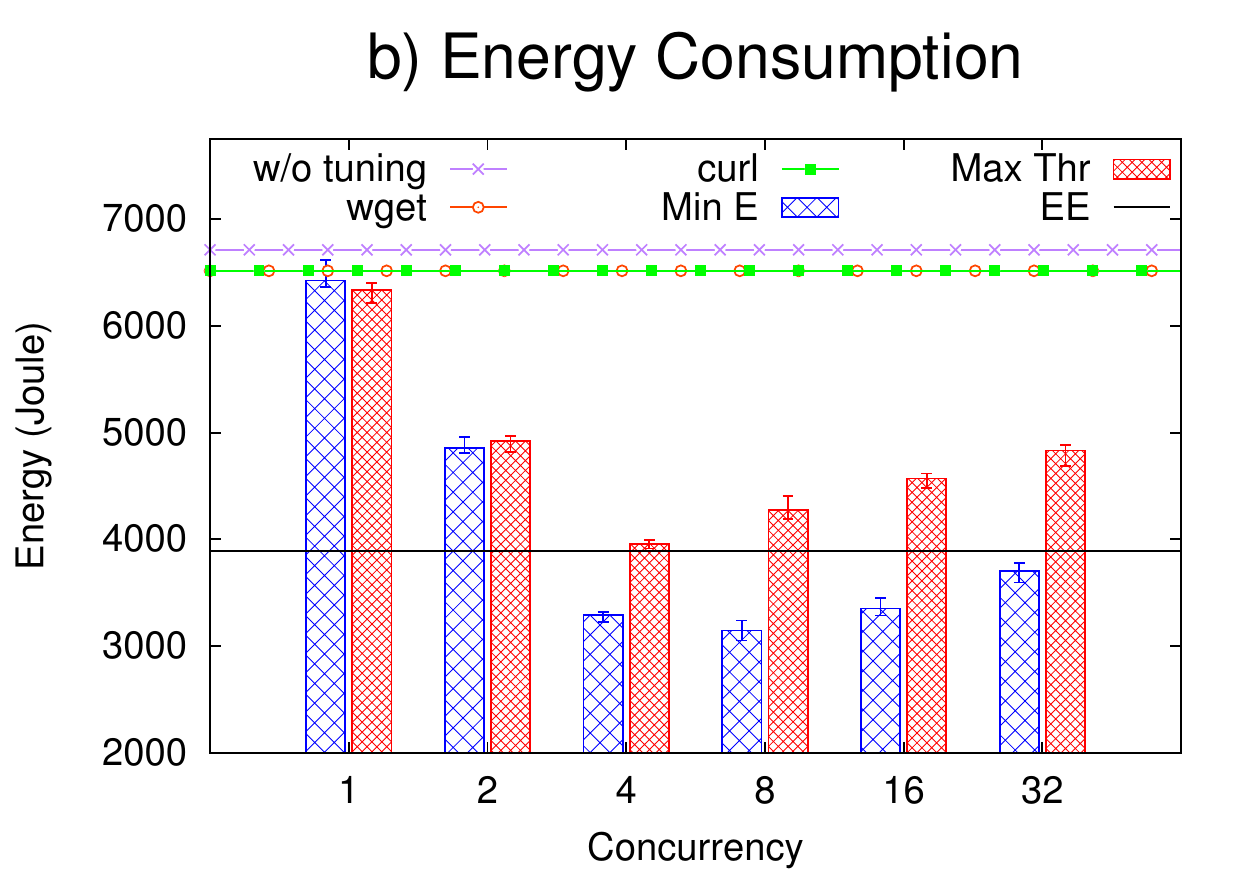}
\includegraphics[keepaspectratio=true,angle=0,width=55mm] {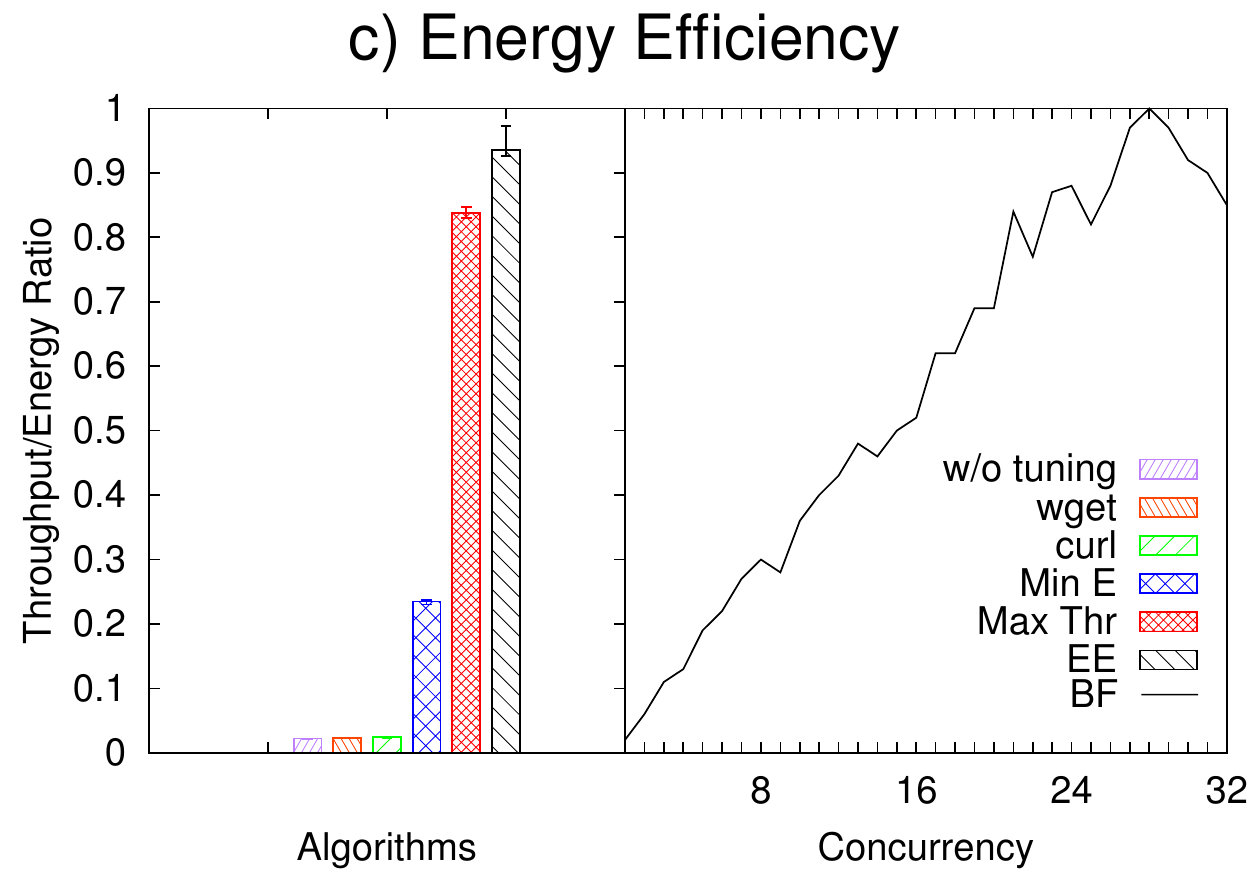}
\end{tabular}
\caption{HTTP transfers from a web server in Austin to a client in Chicago.}  
\label{fig:chamelon10G}
\end{centering}
\vspace{-12 pt}
\end{figure*}

\begin{figure*}[h]
\begin{centering}
\begin{tabular}{cc}
\includegraphics[keepaspectratio=true,angle=0,width=55mm] {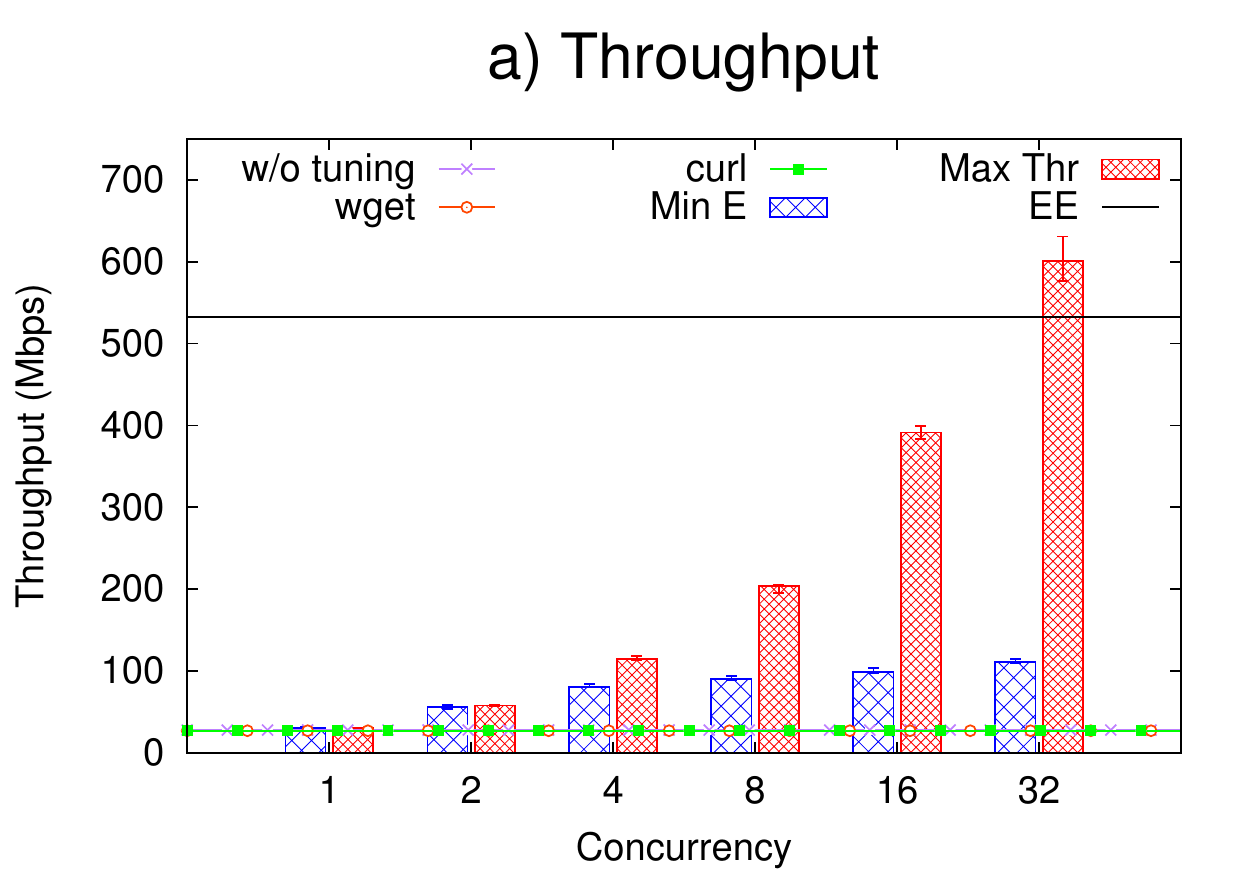}
\includegraphics[keepaspectratio=true,angle=0,width=55mm] {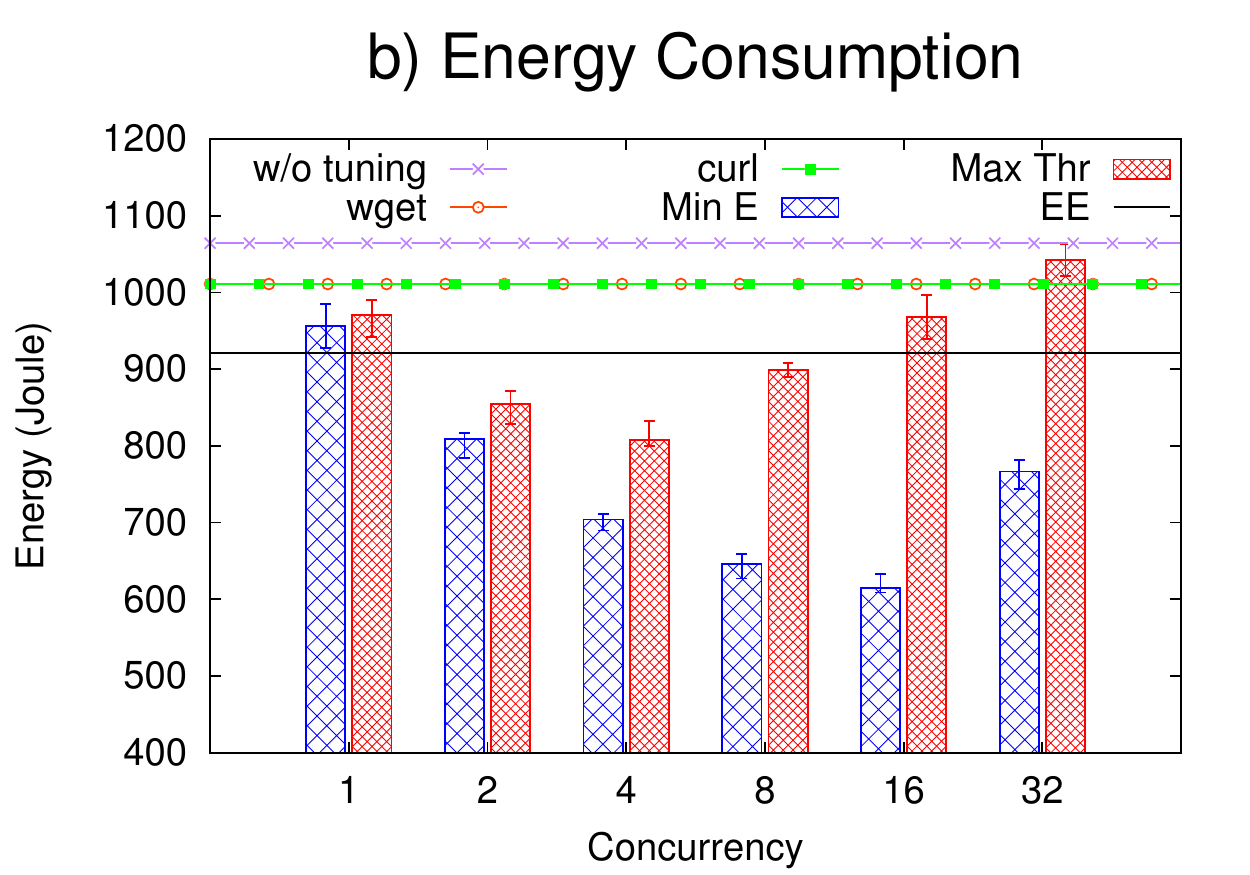}
\includegraphics[keepaspectratio=true,angle=0,width=55mm] {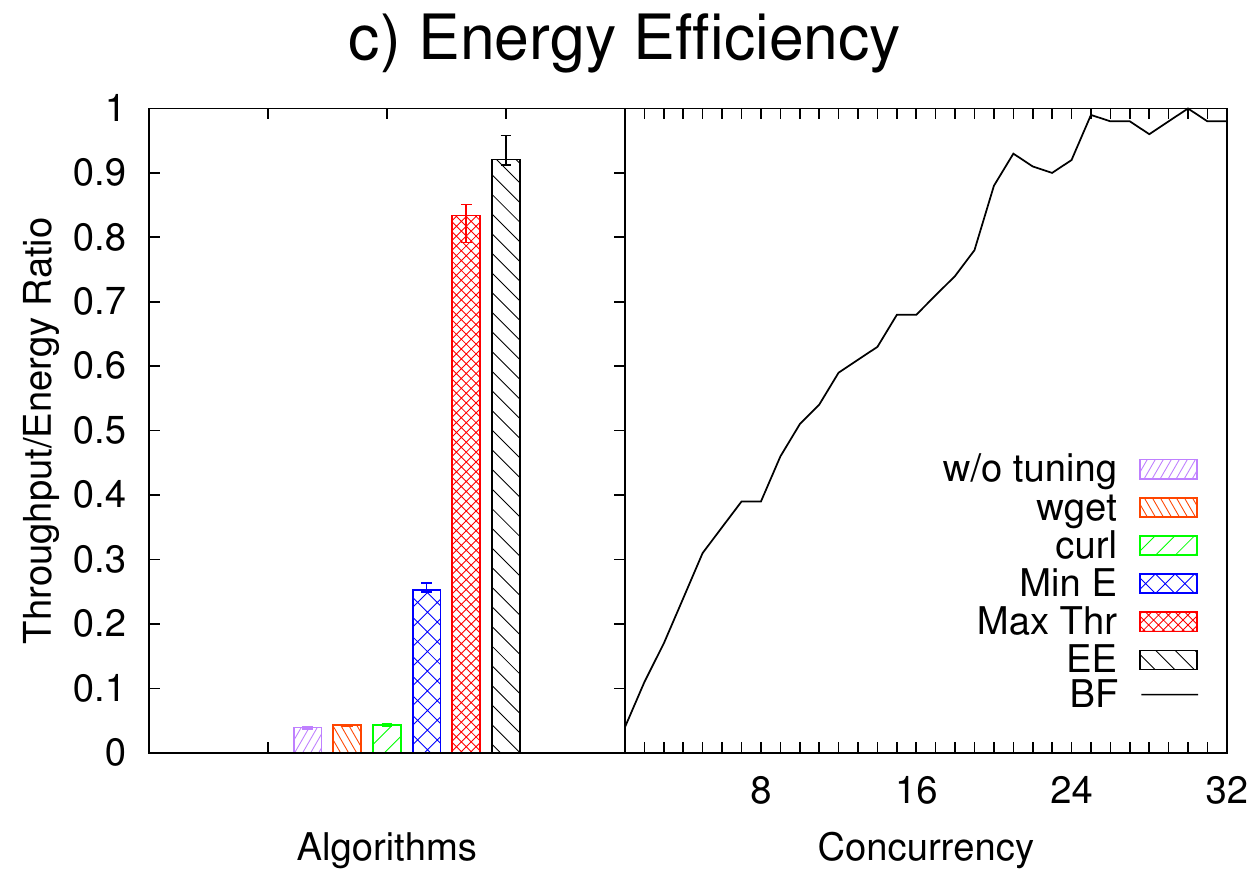}
\end{tabular}
\caption{HTTP transfers from a web server in Virginia to a client in Frankfurt.}
\label{fig:aws}
\end{centering}
\vspace{-12 pt}
\end{figure*}

\section{Experimental Results}

\begin{figure*}[h]
\begin{centering}
\begin{tabular}{cc}
\includegraphics[keepaspectratio=true,angle=0,width=55mm] {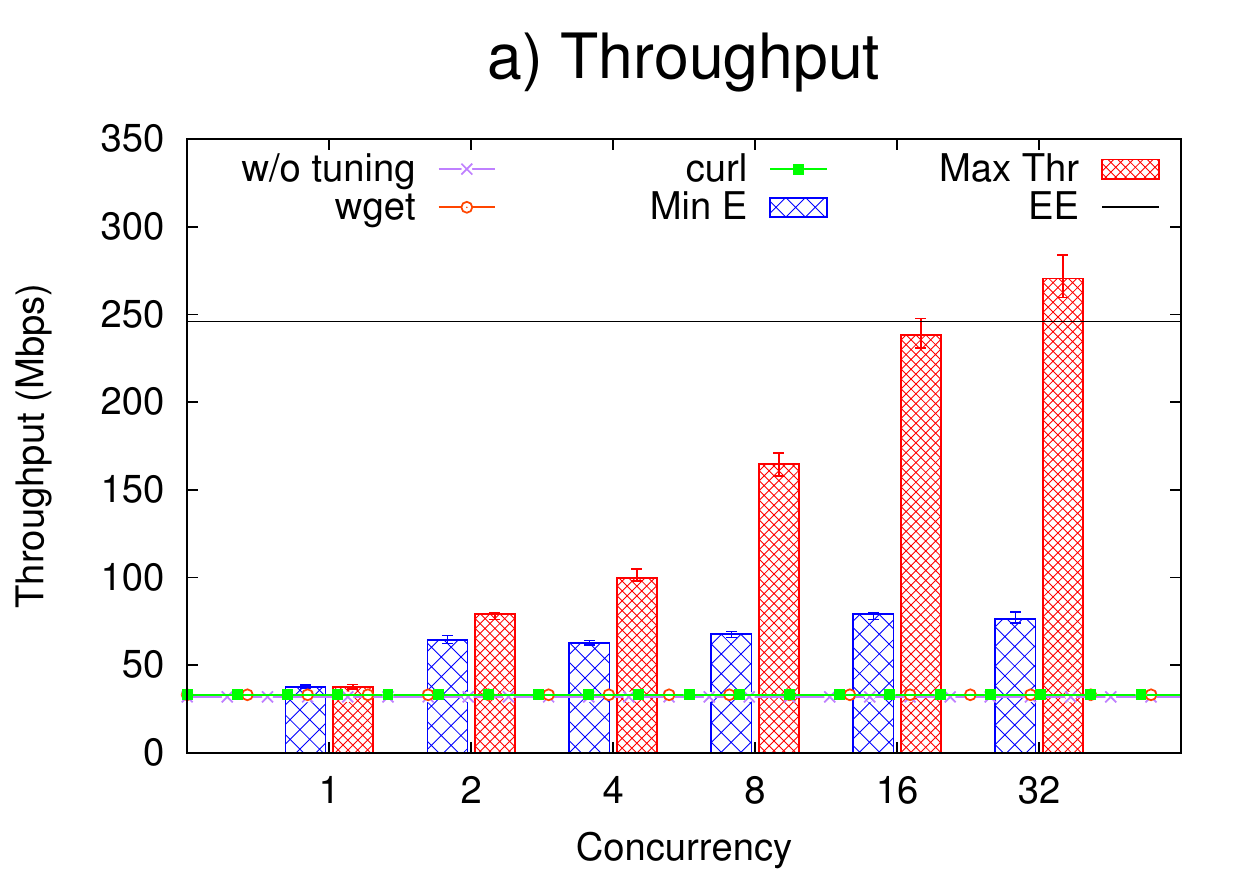}
\includegraphics[keepaspectratio=true,angle=0,width=55mm] {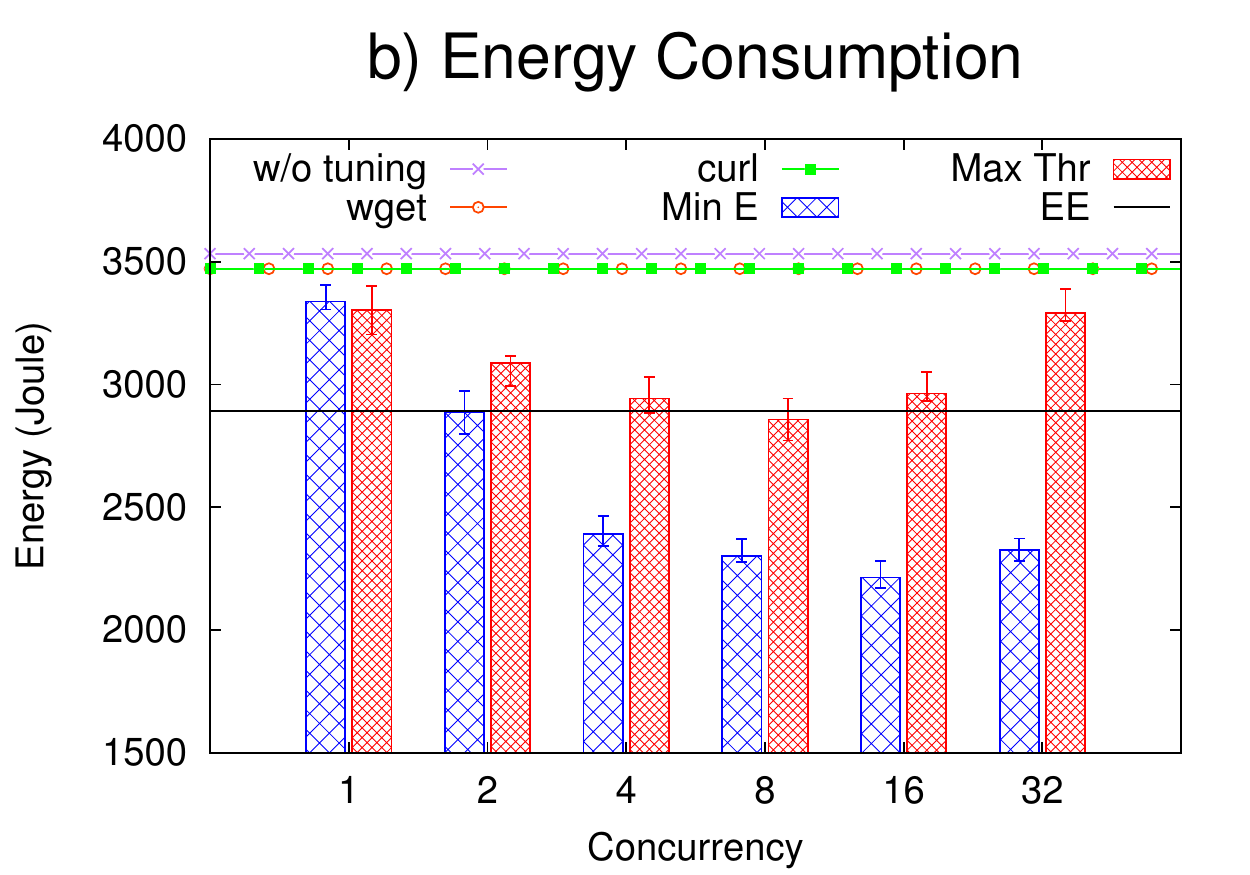}
\includegraphics[keepaspectratio=true,angle=0,width=55mm] {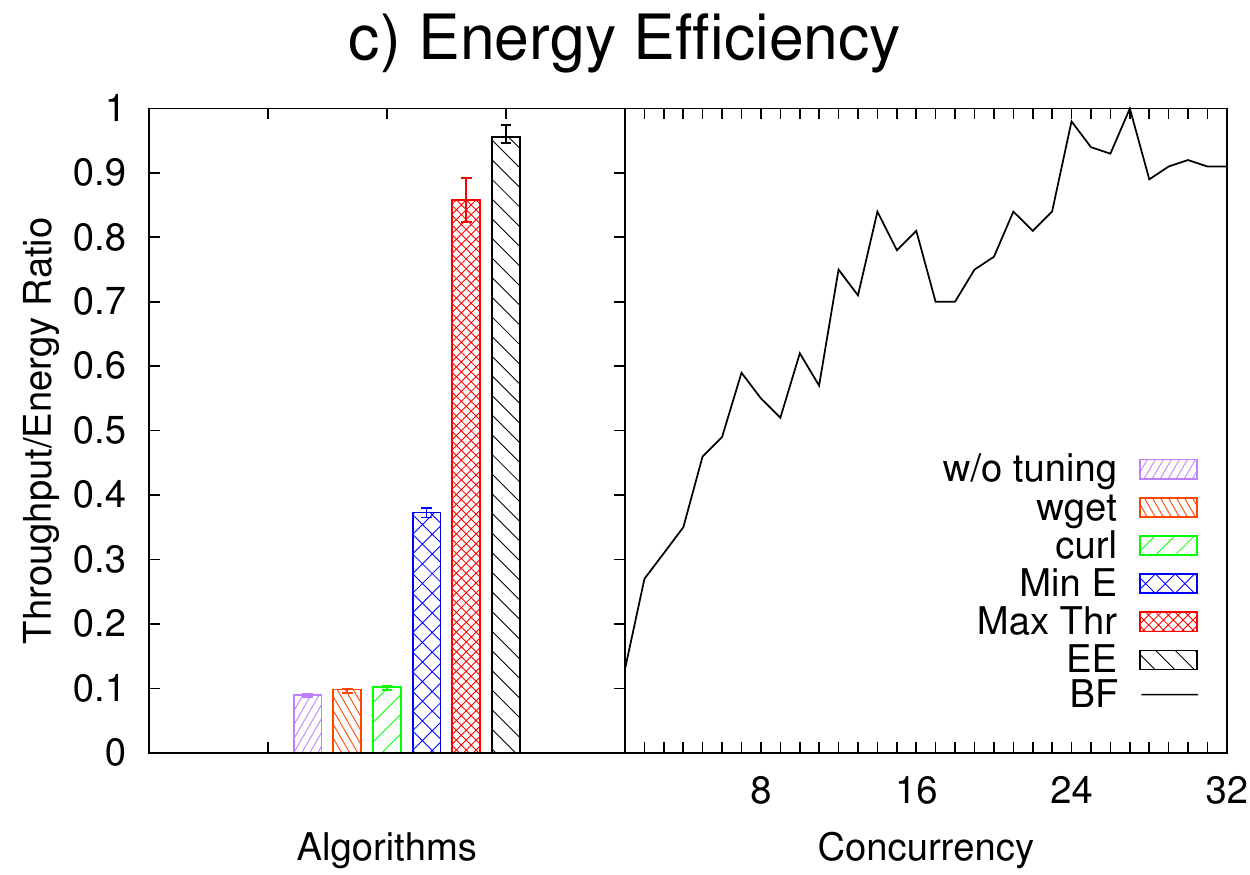}
\end{tabular}
\caption{HTTP transfers from a web server in Austin to a client in Buffalo.}
\label{fig:didclab}
\end{centering}
\vspace{-12 pt}
\end{figure*}

\begin{figure*}[h]
\begin{centering}
\begin{tabular}{cc}
\includegraphics[keepaspectratio=true,angle=0,width=55mm] {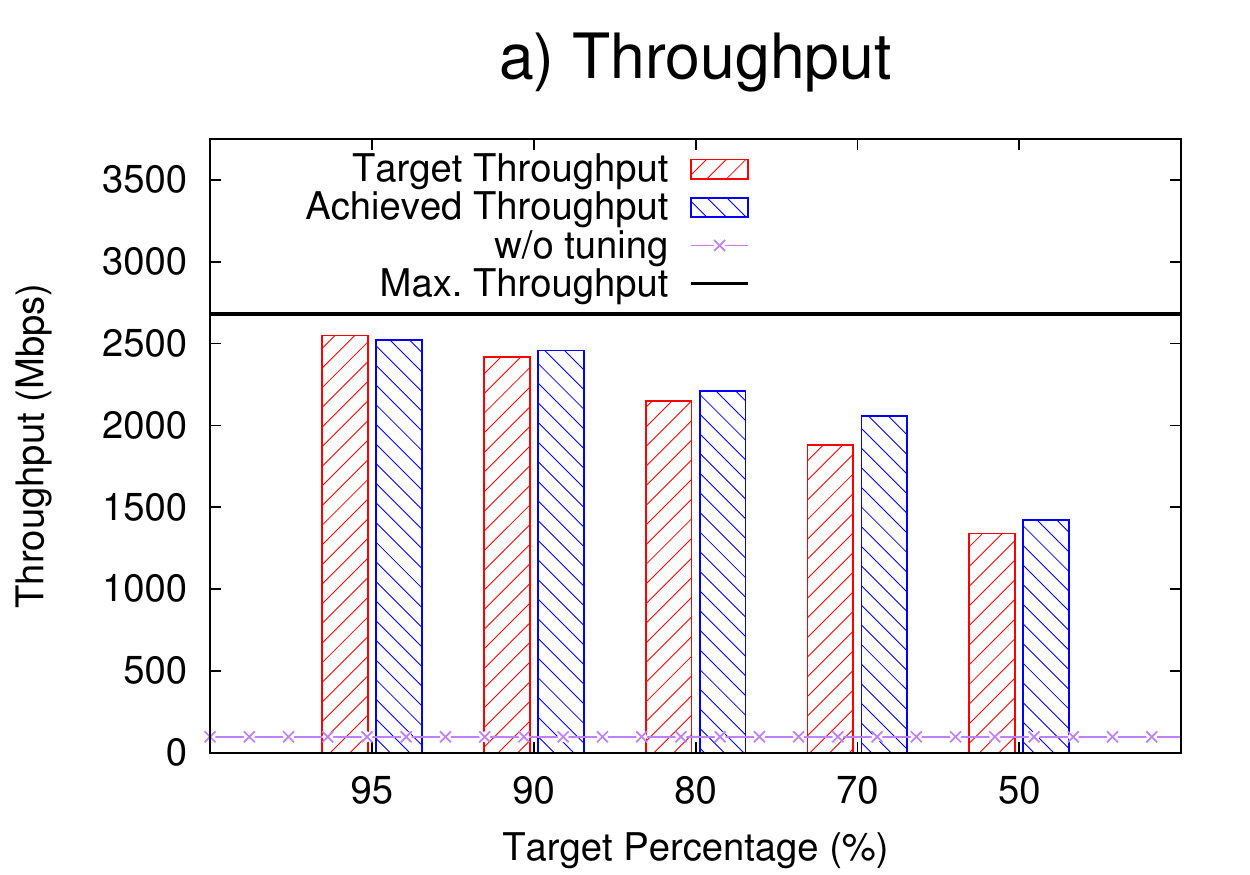}
\includegraphics[keepaspectratio=true,angle=0,width=55mm] {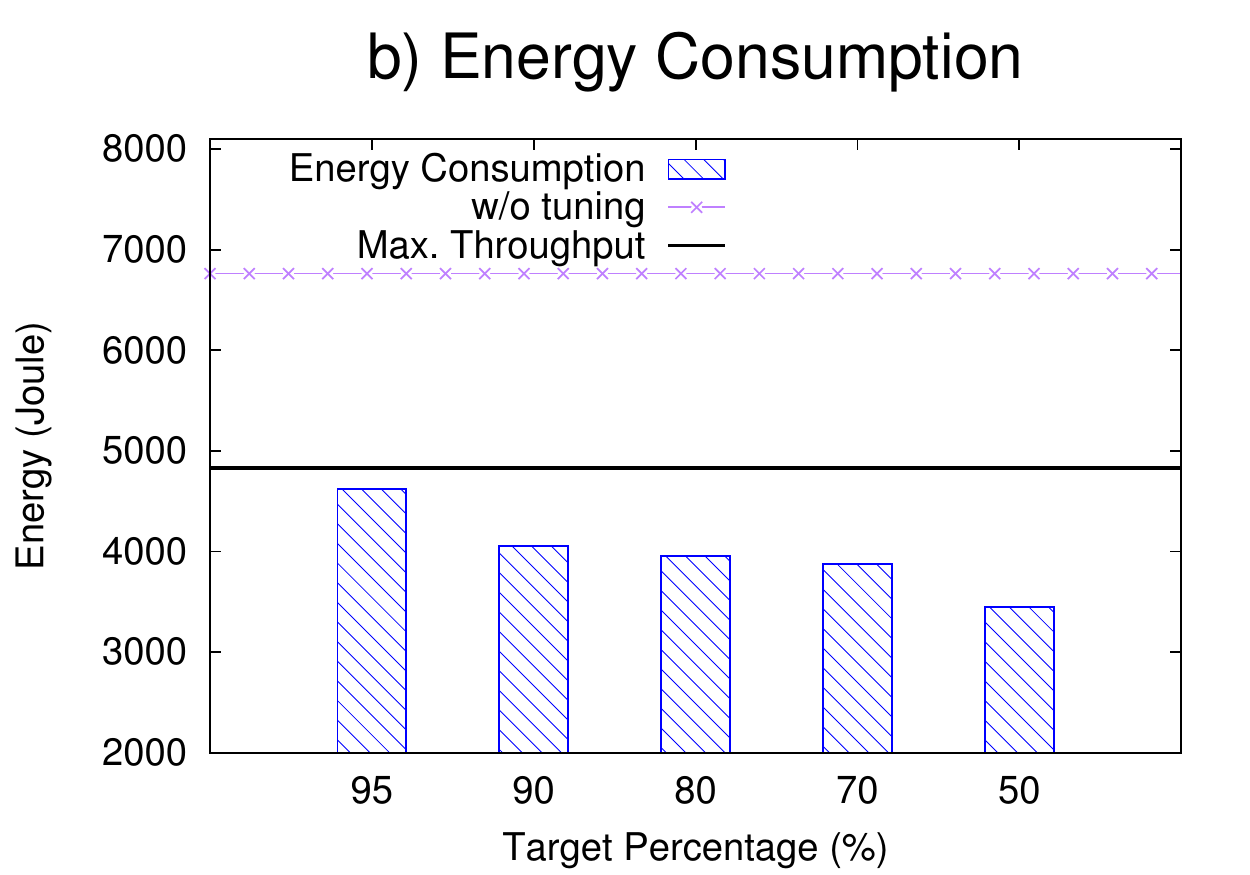}
\includegraphics[keepaspectratio=true,angle=0,width=55mm] {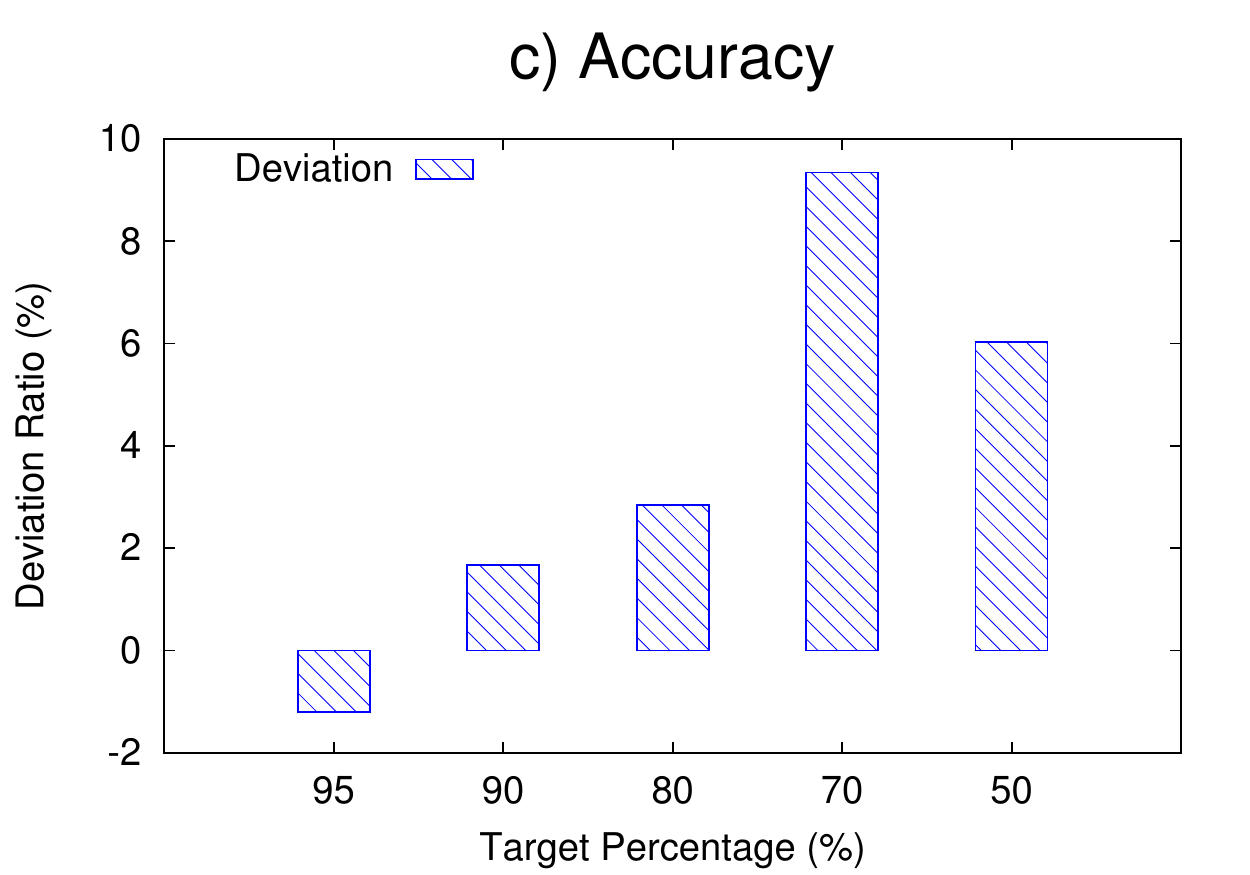}
\end{tabular}
\caption{Flexible throughput experiments from a web server in Austin to a client in Chicago.}
\label{fig:chamelon10G_sla}
\end{centering}
\vspace{-12 pt}
\end{figure*}

Similar to the analysis of protocol parameter experiments, the developed algorithms were tested on the Chameleon Cloud~\cite{chamelon}, DIDCLab~\cite{DIDCLab} and AWS EC2~\cite{ec2} instances as presented in Figure~\ref{fig:netmap}. In the experiments, we used two different datasets due to the different capacities of the utilized networks. For the 10 Gbps links, a 32 GB dataset was used; and for the 1 Gbps links, a 7 GB dataset was used. For both datasets, the file sizes range between 150 KB and 250 MB. The same Apache HTTP server and the custom HTTP client configurations were used during the experiments. We compared the performance of our algorithms with energy-agnostic wget and curl clients as well as our Apache web client at default setting (without any tuning).

We evaluated the performance of MinE and MaxThr algorithms at different concurrency levels defined by the user. On the other hand, EE algorithm takes upper bound for the concurrency level but finds the optimal level itself during its search phase. Since wget and curl tools do not allow to open multiple channels, their performance are independent of the user-defined maximum level of concurrency. For our custom HTTP client, we test it on default (w/o tuning) mode and consider it as a base performance level for a given data transfer. Thus, its performance is also independent of the concurrency level. 

Figure~\ref{fig:chamelon10G} shows the results of the algorithms running on the Chameleon Cloud testbed. The web server is located at an instance in Austin, TX and the web client is located at an instance in Chicago, IL. The network bandwidth between the server and client is 10 Gbps, and the round trip time is 40 ms. The algorithms are compared based on their achieved throughput, energy consumption, and energy efficiency numbers. In accordance with their objective, MaxThr always achieves highest throughput and MinE achieves lowest energy consumption almost at all concurrency levels. Interestingly, while the performance of MinE is five times greater than wget, curl, and w/o tuning mode at concurrency level 8, it also consumes around 60\% less energy than these tools as shown in Figure~\ref{fig:chamelon10G}. This strengthens the idea that in some cases the energy consumption can be decreased and the data transfer throughput can be increased at the same time. Even in the concurrency level 1, our algorithms perform better than other tools. The achievement stems from using optimal pipelining and parallelism values starting from initial configuration. It also shows that even concurrency is the most influential parameter, parallelism and pipelining can still make a difference in the throughput and energy consumption, thus should be tuned carefully. \EE~aims to find the sweet spot where the throughput/energy (energy efficiency) ratio is maximized. It searches the concurrency level in the search space, bounded by 1 and {\em maxChannels}, after which the throughput increase is surpassed by the energy consumption increase. As given in Algorithm~\ref{alg:EE}, it evaluates  the throughput/energy ratio of multiple concurrency levels and picks the concurrency level with the highest throughput/energy ratio and runs the rest of the transfer with it. Compared to MaxThr, \EE~consumes 25\% less energy in trade off 10\% less throughput for concurrency level 32 at which MaxThr achieves the highest throughput. At concurrency level 4, MaxThr consumes similar amount of energy while obtaining 4 times less throughput compared to \EE~which can justify the argument of consuming less amount of energy without sacrificing the data transfer throughput.

In order to compare the performance of the \EE~algorithm to the ideal case, we ran a brute-force search (BF) algorithm to find the concurrency level which maximizes the throughput/energy ratio. BF is a revised version of the \EE~algorithm in a way that it skips the search phase and runs the transfer with pre-defined concurrency levels. BF search range is bounded by 32 since the throughput/energy ratio follows a steadily decreasing pattern after around a value of 28 as shown in Figure~\ref{fig:chamelon10G} (c). The concurrency level which yields the highest throughput/energy ratio is considered as the best possible value under these conditions, and the throughput/energy ratio of all other algorithms are compared to this value. As a result, we observe that the concurrency level chosen by \EE~can yield as much as 94\% throughput/energy efficiency compared to the best possible value obtained by BF. On the other hand, while MinE is successful in consuming the least amount of energy for each concurrency level, it can only reach around 25\% of the best possible throughput/energy ratio. 

We also tested our algorithms using Amazon's EC2 nodes located in Frankfurt and Virginia. Between these two nodes, the network bandwidth is 1 Gbps, and the round trip time is 90 ms. The wget, curl, and w/o tuning mode again yield the lowest throughput and consume most energy due to lack of parameter tuning. Additionally, MaxThr achieves highest throughput and MinE achieves lowest energy consumption almost at all concurrency levels. MinE consumes minimum energy at concurrency level 16 and decreases energy consumption 75\% while increasing throughput three times with respect to the base level. Although, the energy consumption of MaxThr and EE are close to each other at concurrency level 8, the throughput values differ considerably and EE achieves 2 times more throughput compared to MaxThr. While MaxThr's throughput increases as the concurrency level increases, its power consumption curve follows a parabolic pattern and reaches the minimum point at concurrency level 4. One of the reasons for this observed pattern is that the transfer instances used on AWS have four cores and energy consumption per core decreases as the number of active cores increases~\cite{alan2014energy1}. When concurrency level goes above 4, then each core starts running more number of threads and this leads to an increase in energy consumption per core. When we look at the energy efficiency graph, EE algorithm can reach as much as 92\% of the optimal value while MaxThr reaches 83\%, MinE reaches 26\%, and other tools can not exceed 10\% level.

\begin{figure*}[t]
\begin{centering}
\begin{tabular}{cc}
\includegraphics[keepaspectratio=true,angle=0,width=55mm] {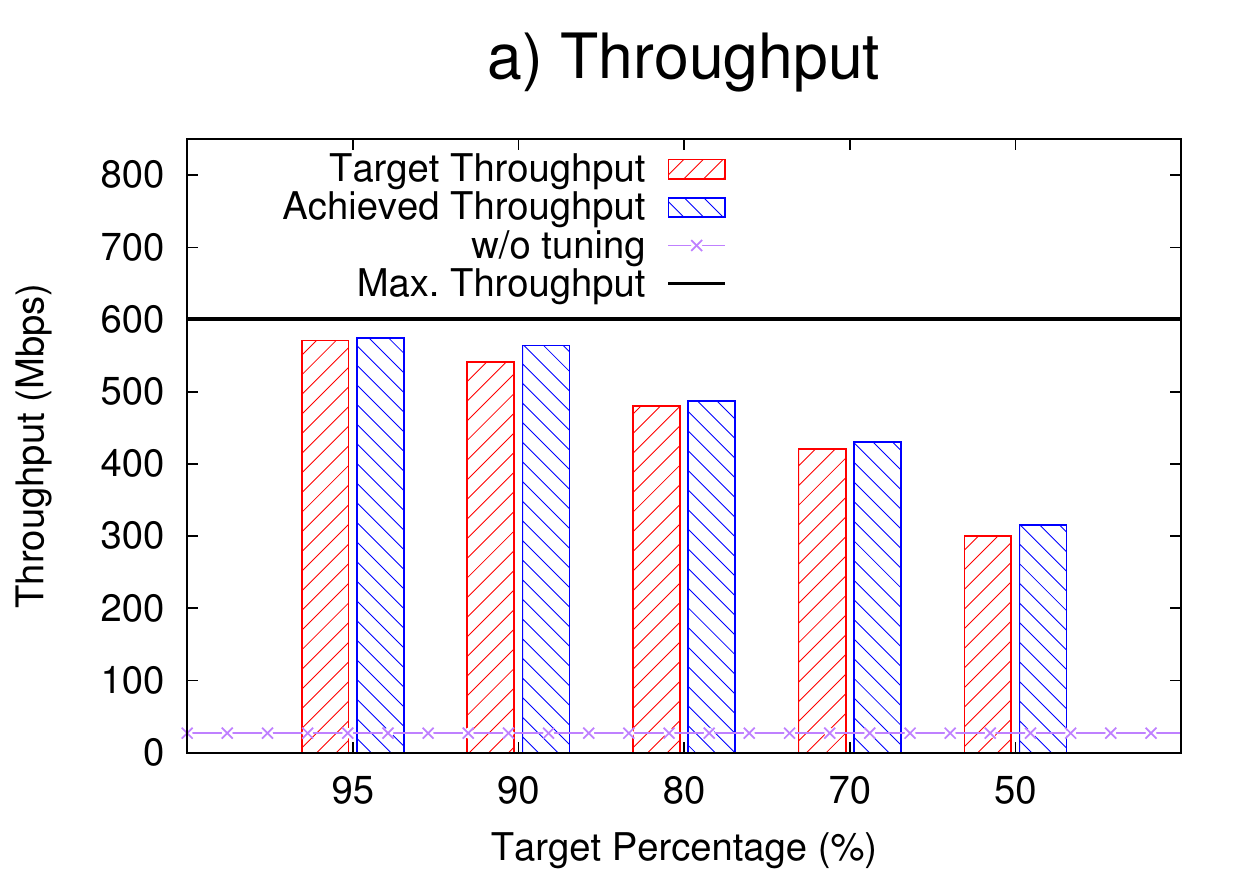}
\includegraphics[keepaspectratio=true,angle=0,width=55mm] {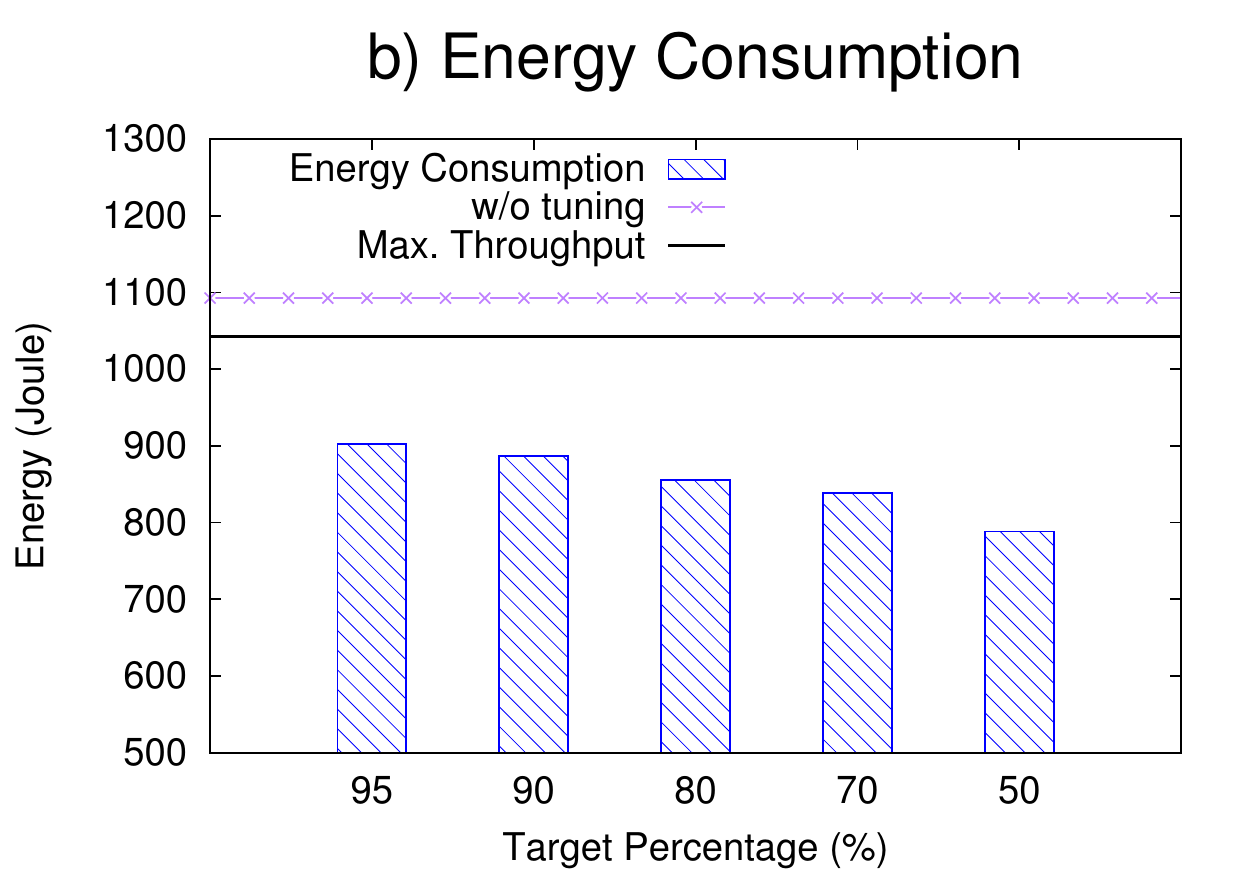}
\includegraphics[keepaspectratio=true,angle=0,width=55mm] {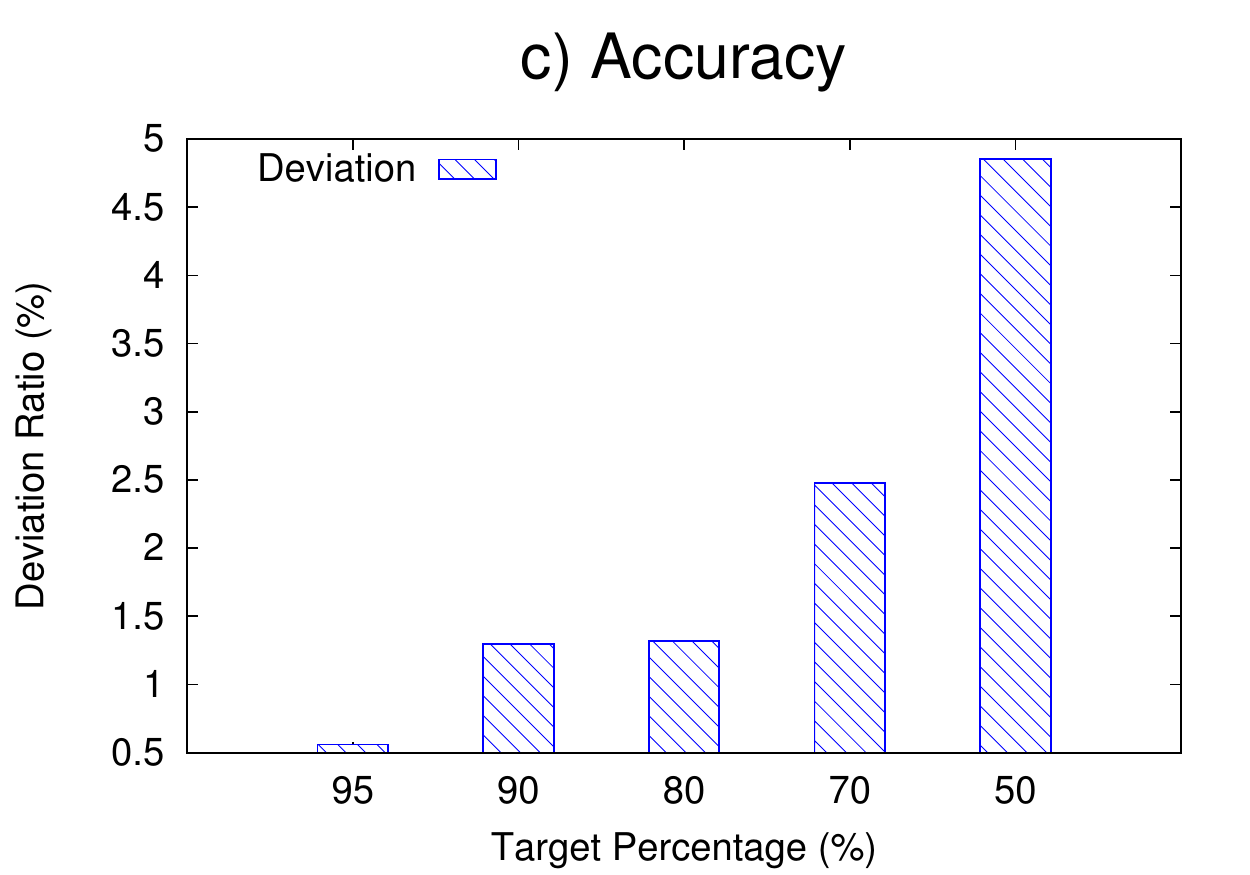}
\end{tabular}
\caption{Flexible throughput experiments from a web server in Virginia to a client in Frankfurt.}
\label{fig:aws_sla}
\end{centering}
\vspace{-12 pt}
\end{figure*}

\begin{figure*}[h]
\begin{centering}
\begin{tabular}{cc}
\includegraphics[keepaspectratio=true,angle=0,width=55mm] {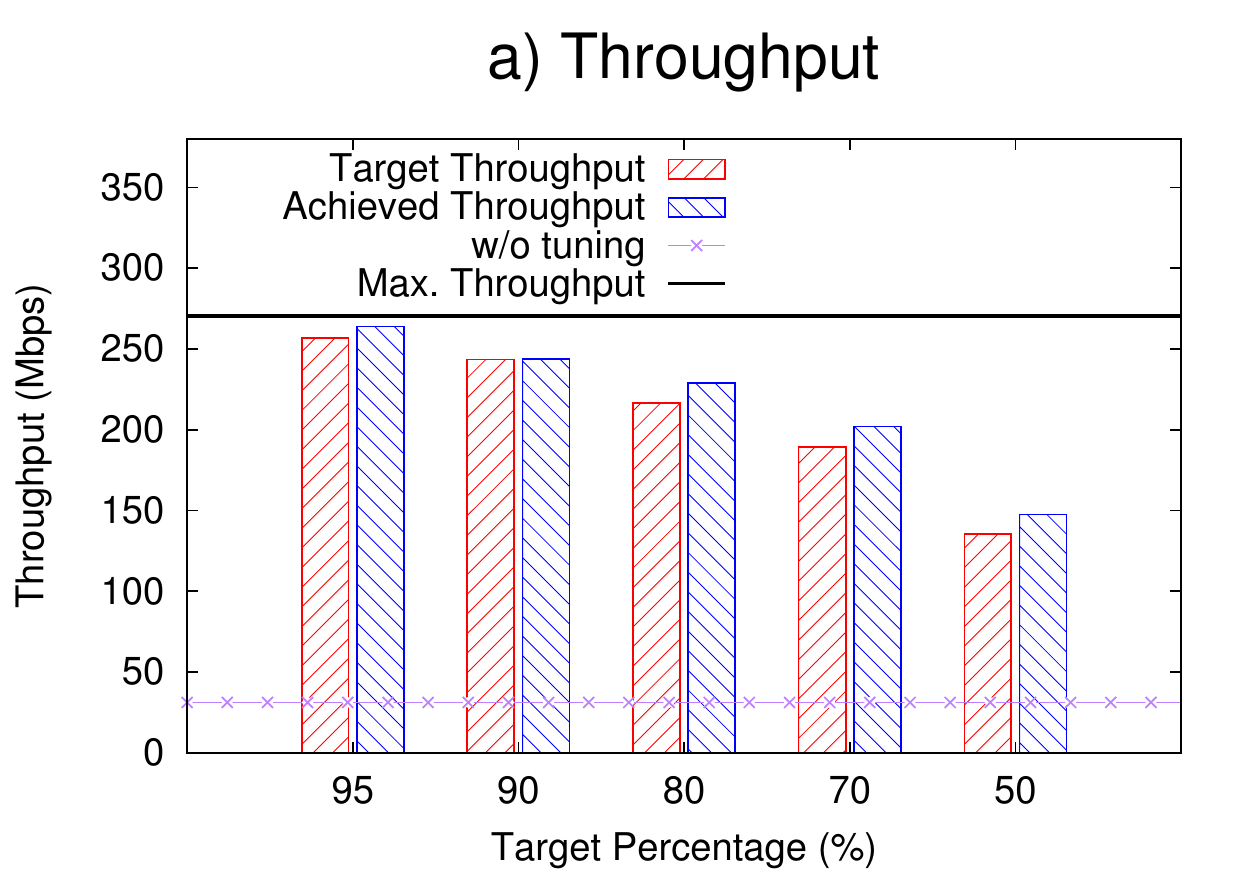}
\includegraphics[keepaspectratio=true,angle=0,width=55mm] {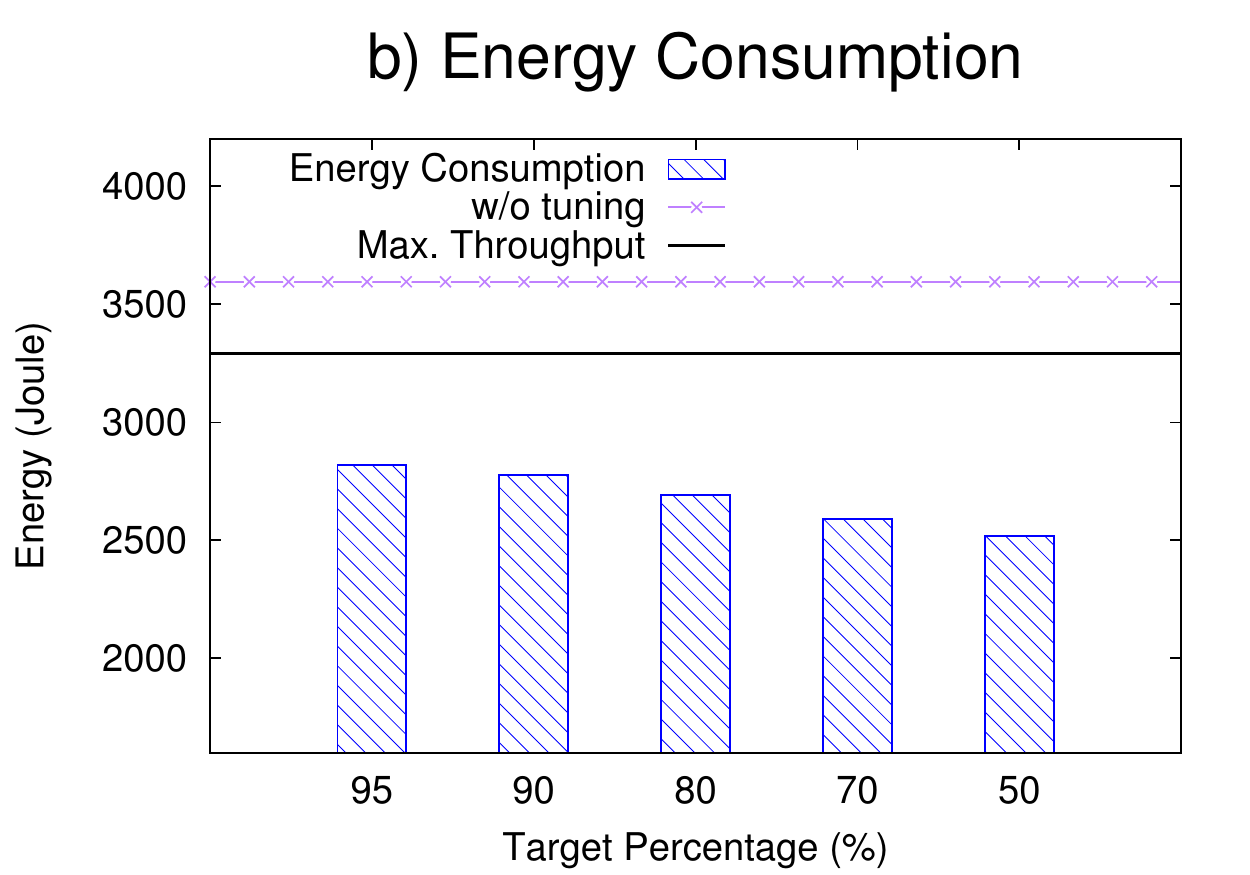}
\includegraphics[keepaspectratio=true,angle=0,width=55mm] {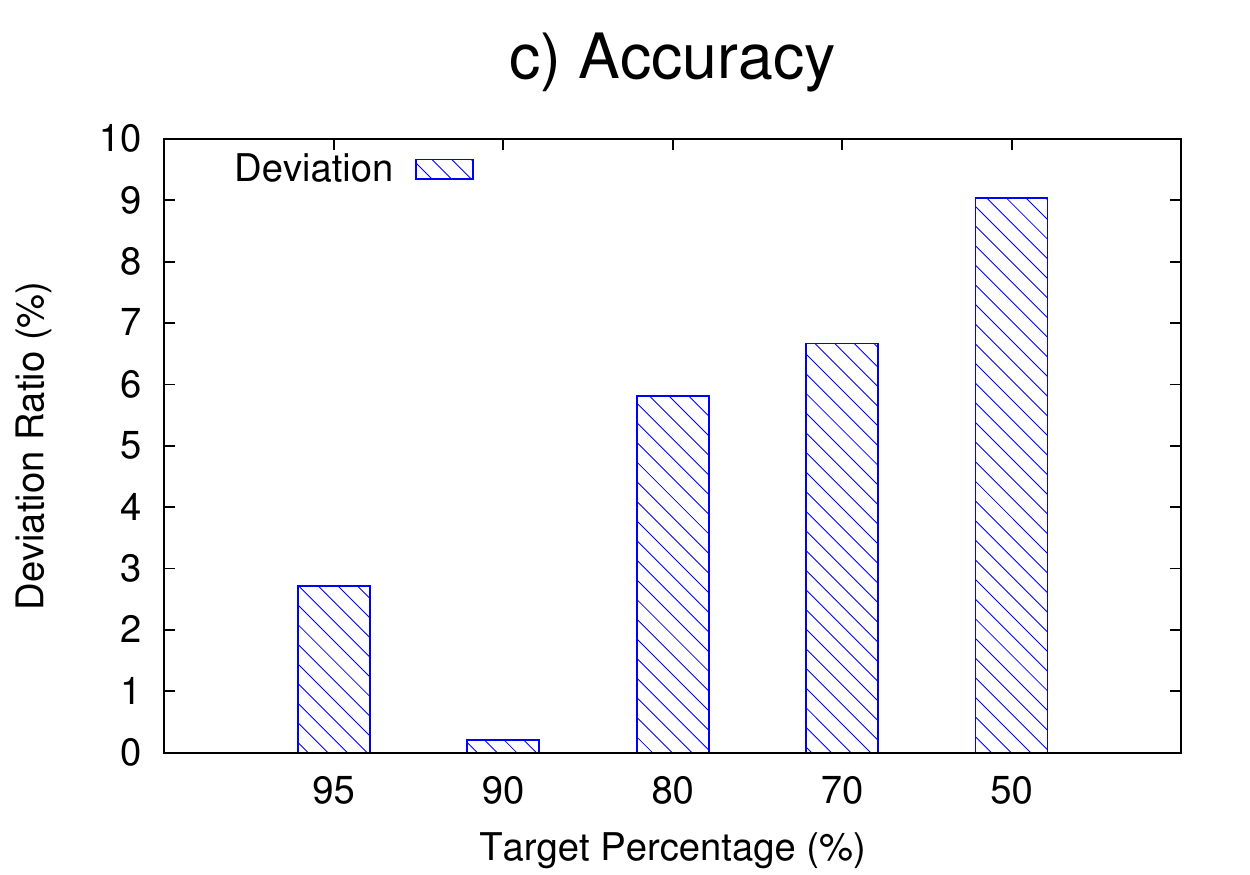}
\end{tabular}
\caption{Flexible throughput experiments from a web server in Austin to a client in Buffalo.}
\label{fig:didclab_sla}
\end{centering}
\vspace{-12 pt}
\end{figure*}

Finally, we evaluated our SLA based algorithms between a web server at the Chameleon Cloud node in Austin,TX and the web client at DIDCLab in Buffalo,NY as presented in Figure~\ref{fig:didclab}. Similar to the previous environments, MinE and MaxThr realize their goals. MaxThr algorithm increases throughput from 37 Mbps to 290 Mbps despite the shared low-bandwidth internet connection between the server and client; and MinE algorithm decreases energy consumption 85\% with respect to the base power consumption. Unlike the other experiments, MinE and MaxThr achieve their best throughput/energy ratio at comparatively low concurrency levels in this setting. This strengthens the motivation behind \EE~as it is designed to capture the network-specific sweet spots in which throughput/energy ratio is maximized.

We tested the FlexibleThr algorithm with different levels of throughput targets proportioned to the maximum throughput achieved by the MaxThr algorithm in this specific environment. In FlexibleThr, x\% target percentage means that the algorithm tries to achieve a transfer throughput more than x\% of the maximum throughput possible (i.e. with at most 100\% -- x\% performance loss). For example, maximum throughput achieved by MaxThr in Chameleon Cloud network with 10G connection is around 2600 Mbps, so 95\% target percentage corresponds to 2470 Mbps or more transfer rate. 

\FT~is able to deliver all various throughput requests except 95\% target throughput percentage at the Chameleon Cloud with 10G network since \FT~is unable to reach the requested throughput on this network even after reaching the maximum level of concurrency. As presented in Figure~\ref{fig:chamelon10G}(b),~\ref{fig:aws}(b) and ~\ref{fig:didclab}(b) after some specific concurrency level, energy consumption starts to increase. Thus, achieving target throughput with minimum possible concurrency level will minimize energy consumption. \FT~is able to achieve all throughput expectations within 10\% deviation rate as shown in Figure~\ref{fig:chamelon10G_sla}(c),~\ref{fig:aws_sla}© and ~\ref{fig:didclab_sla}(c). Since the effect of each new channel creation on throughput is higher when the total number of channels is small, accuracy decreases as \FT~is expected to provide low transfer throughput. Figure~\ref{fig:chamelon10G}(b),~\ref{fig:aws}(b) and ~\ref{fig:didclab}(b) depicts energy consumption comparison between MaxThr, w/o tuning mode, and \FT~algorithm. \FT~can deliver requested throughput while decreasing the energy consumption by up to 50\% for Chameleon Cloud with 10G interconnection, up to 38\% for AWS transfers and up to 42\% Chameleon Cloud and DIDCLab transfers respectively. Finally, if customers are flexible in transferring their data with some reasonable delay, \FT~algorithm helps the service providers to save from the energy consumption considerably and also the service providers can possibly offer low-cost and flexible data transfer options to their customers in return of delayed transfers.

\section{Conclusion}

In this paper, we analyzed different application-layer parameters that affect both the throughput and the energy consumption of HTTP, which is the de-facto transport protocol for Web services. Different parameters such as pipelining, parallelism and concurrency levels play an important role in the achievable network throughput and the total energy consumption. We introduced SLA-based algorithms which can decide the best combination of these parameters based on user-set energy efficiency and performance criteria. Our experimental results show that significant amount of energy savings (up to 80\%) can be achieved at the sending and receiving nodes during data transfers with no or minimal performance penalty if the correct parameter combination is used. In some cases, 
both the throughput can be maximized and the energy consumption can be minimized at the same time. With the help of our SLA-based energy-efficient data transfer algorithms, the Internet service providers will be able to minimize the energy consumption during data transfers without compromising the SLA with the customer in terms of the promised performance level, but still execute the transfers with minimal energy levels given the requirements. 
%

%
%
\bibliographystyle{plain}
\bibliography{main}
\end{document}